\documentclass[preprint,a4paper,12pt]{article}      
\usepackage{fullpage}
\usepackage{graphicx}
\usepackage{natbib}
\usepackage{amsmath}
\usepackage{bm}
\usepackage{booktabs}
\usepackage{url}
\usepackage[caption=false]{subfig}
\usepackage{color}
\usepackage{multirow}
\usepackage{lineno}

\begin{document}
\begin{center}
		{\Large A Bayesian semi-parametric hybrid model for spatial extremes with unknown dependence structure}\\\vspace{6pt}
		{\large Yuan Tian\footnote[1]{North Carolina State University} and Brian J. Reich$^1$}\\
\end{center}

\begin{abstract}
	The max-stable process is an asymptotically justified model for spatial extremes. In particular, we focus on the hierarchical extreme-value process (HEVP), which is a particular max-stable process that is conducive to Bayesian computing. The HEVP and all max-stable process models are parametric and impose strong assumptions including that all marginal distributions belong to the generalized extreme value family and that nearby sites are asymptotically dependent. We generalize the HEVP by relaxing these assumptions to provide a wider class of marginal distributions via a Dirichlet process prior for the spatial random effects distribution. In addition, we present a hybrid max-mixture model that combines the strengths of the parametric and semi-parametric models. We show that this versatile max-mixture model 
	accommodates both asymptotic independence and dependence and can be fit using standard Markov chain Monte Carlo algorithms. The utility of our model is evaluated in Monte Carlo simulation studies and application to Netherlands wind gust data.
	
\end{abstract}

\section{Introduction}
\label{s:intro}
Extreme value analysis plays a significant role in climate research. Extreme events such as unusually high temperature, major precipitation and life threatening hurricanes occur with small probability but may have catastrophic consequences. It is therefore of great significance to make inferences and predictions about these rare events. As the variables of interest in climate research are typically recorded over space, spatial extreme models have proven to be useful tools \citep{doi:10.1146/annurev-statistics-010814-020133}. 

One common approach is to model these rare events by the max-stable process. Max-stable processes are natural extensions of multivariate extreme value distributions to infinite dimensions. They arise as the limits of maxima of independent copies of stochastic processes. 

Based on their spectral representation \citep{de1984spectral}, a series of parametric models have been proposed \citep{brown1977extreme, opitz2013extremal,schlather2002models,smith1990max}. However, the model fitting is computationally difficult as the closed form of the full likelihood involves a Bell number of terms,
which increases rapidly as the number of spatial locations increases.
Existing approaches to parameter estimation include methods based on composite likelihood \citep{huser2013composite,ribatet2012bayesian}, exact likelihood with additional information \citep{stephenson2005exploiting,thibaud2015efficient, wadsworth2013efficient} and M-estimators \citep{einmahl2016m,einmahl2012m,yuen2014crps}. In addition, the prediction of spatial extremes for an unobserved location is also difficult and only recently addressed by \cite{dombry2012conditional, dey2016conditional, dombry2013regular}.

Unlike the likelihood based methods, \cite{reich2012hierarchical} proposed a Bayesian hierarchical extreme value process (HEVP), which is a particular max-stable process. This model can be easily implemented via MCMC algorithm. In addition, prediction is also straightforward by sampling from the posterior predictive distribution. 

Despite the desirable properties, the HEVP model is fairly restricted in practice.
First, the HEVP model assumes that all marginal distributions follow the GEV distribution. However, the marginal distributions of environmental processes are not necessarily GEV distributed. It is therefore appealing to propose a model that can span a wider family of spatial processes to allow for more flexibility while still being stochastically centered on the asymptotic GEV distribution.
Second, the HEVP model imposes strong asymptotic dependence between two nearby sites. 
This assumption may also be violated.
For example, many complex environmental processes exhibit weakening tail dependence as events become more extreme. 
One way to address this is to specify the dependence structure in advance and choose different models accordingly. However, it is challenging in practice because asymptotic dependence is generally difficult to infer.
The recent literature has considered spatial models encompassing
both asymptotic dependent and independent classes. \cite{coles2002models} proposed a model based on powers of survival functions. The max-mixture model proposed by \cite{wadsworth2012dependence} is a hybrid of the asymptotically dependence and asymptotically independence sub-families. \cite{huser2017bridging} proposed a Gaussian scale mixture model that allows for both asymptotic dependence structures.
\cite{bopp2018hierarchical} extended the HEVP model to a max-infinitely divisible process model that can cope with flexible tail behaviors. {\cite{hazra2018semiparametric} proposed a nonparametric Bayesian model centered on a skewed-$t$ process for spatial extremes.}
However, in the aforementioned models, one sub-family only occurs at the boundary points, thus makes the inference non-regular. Recently, a new model proposed in \cite{huser2018modeling} achieved a smooth transition between the two dependence structures. 

In this paper, we first extend the current HEVP model to nonparametric Bayesian model with stick-breaking prior (SB). The stick-breaking prior relaxes parametric assumptions and allows for more flexibility by expanding the class of marginal distributions. Unlike the HEVP model, the SB model is asymptotically independent. Therefore, to borrow strength from both the HEVP and the SB models, we further propose a hybrid max-mixture (MM) model which is based on maximum of powers of an asymptotically dependent process and an asymptotically independent  process.  
The MM model nests the HEVP and the SB models as special cases
so when the parametric model is correctly specified, the MM model will reduce to the HEVP model and when the data dictate that the parametric model is wrong, the MM model is a more flexible model and not based on a particular set of asymptotic assumptions. 
We show that the MM model can accommodate unknown dependence structures including both asymptotically dependent and independent subfamilies and achieves a smooth transition between these 
two subfamilies.
In our Bayesian analysis we obtain the posterior probability that the process is asymptotically dependent, which provides important insights about the extreme value process under consideration.
In this way the posterior predictive distribution of the MM model naturally averages over uncertainty in the dependence class.
Both extensions (SB and MM) inherit 
the computational advantages of the HEVP model as they
can be fit using standard MCMC algorithms and can deliver predictions at unobserved locations.

The remainder of the paper proceeds as follows. Sections \ref{s:method} introduce the statistical methods and corresponding theoretical properties. The proposed methods are evaluated using a simulation study in Section \ref{s:sim} and applied to the motivating data sets in Section \ref{s:app}. We conclude the paper with a discussion section in Section \ref{s:discussion}.

\section{Statistical methods}\label{s:method}
\subsection{Review of the hierarchical extreme value process}\label{s1:review}
Let $Y_t(s)$ be the response at time $t$ and location $s$. Assuming $Y_t(\cdot)$ follows a spatial max-stable process for time $t$, then $Y_t(s)$ marginally follows the 
Generalized Extreme Value (GEV) distribution with location $\mu_t(s)$, scale $\sigma_t(s)$ and shape $\xi_t(s)$. For simplicity, we assume these GEV parameters are constant over time so that the marginal distribution is $Y_t(s) \sim \mathrm{GEV}\{\mu(s),\sigma(s),\xi(s)\}$. Equivalently, $Y_t(s)$ can be expressed as 
\begin{equation}
\setlength{\jot}{10pt}
Y_t(s) = \mu(s)+\frac{\sigma(s)}{\xi(s)}[X_t(s)^{\xi(s)}-1], \nonumber
\end{equation}
where $X_t(\cdot)$ is the residual max-stable process with unit Fr{\'e}chet margins, $X_t(s)\sim \mathrm{GEV}(1,1,1)$. 

Reich and Shaby (2012) propose the hierarchcal extreme value process (HEVP) model for the residual process $X_t(\cdot)$. In the HEVP model, $X_t(s)$ is constructed as the product of two independent processes, $X_t(s)=U_t(s)\theta_t(s)$. The non-spatial term is $U_t(s)\stackrel{i.i.d}{\sim} \mathrm{GEV}(1, \alpha, \alpha)$ for some $\alpha \in (0,1]$. The spatial term is $\theta_t(s)=\{\sum_{l=1}^{L}A_{lt}\omega_l(s)^{1/\alpha}\}^\alpha$, where the kernel basis functions satisfy  $\omega_l(s)\geq 0$ with $\sum_{l=1}^L \omega_l(s) = 1$ for all $s$. The random effects $A_{lt}$ follow the positive stable distribution \citep{fougeres2009models} with density $P(A|\alpha)$ and Laplace transformation $\int_0^{\infty}\exp(-At)P(A|\alpha)dA=\exp(-t^\alpha)$. This model for $X_t(s)$ ensures max-stability and unit Fr{\'e}chet marginal distributions. 

Marginalizing over the $U_t(s)$ gives the hierarchical model
\begin{align}
	\label{RS}
	Y_t(s)|A_{1t}, \ldots, A_{Lt} &\stackrel{i.i.d.}{\sim} \mathrm{GEV}\{\mu_t^\star(s), \sigma_t^\star(s),\xi_t^\star(s)\},\\
	A_{1t}, \ldots, A_{Lt} & \stackrel{i.i.d.}{\sim} \mathrm{PS}(\alpha), \nonumber  
\end{align}
where $\mu_t^\star(s)=\mu(s)+\frac{\sigma(s)}{\xi(s)}[\theta_t(s)^{\xi(s)}-1], \sigma_t^\star(s) = \alpha \sigma(s) \theta_t(s)^{\xi(s)}$ and $\xi_t^\star(s)=\alpha\xi(s)$. Further, marginalizing over the random effects $A_{1t}, \ldots, A_{Lt}$ gives the marginal distribution 
$$Y_t(s) \sim \mathrm{GEV}\{\mu(s), \sigma(s),\xi(s)\}.$$
This model is max-stable and applicable in high-dimensions due to its conditional representation in terms of $L$ positive stable variables. 
Both terms $U(\cdot)$ and $\theta(\cdot)$ share the parameter $\alpha$ that determines the relative magnitude of the non-spatial and spatial effects. For a small $\alpha$, the contribution of the spatial term $\theta(s)$ dominates the non-spatial term $U(s)$ and vice-versa for large $\alpha$. The HEVP model also contains the Smith model \citep{smith1990max} as a limiting case when $\alpha \to 0$. The Smith model is often criticized in practice due to its lack of flexibility \citep{dey2016extreme}, but by incorporating the non-spatial terms $U_t(s)$ the HEVP model introduces more flexibility than the Smith model.

The marginal distribution function of the residual max-stable process $X_t(\cdot)$ at location $s$ is $F_{HEVP}(c;s)=\mathrm{P}\{X_t(s)<c\}=\exp(-1/c)$ and the joint distribution function of the residual max-stable process $X_t(\cdot)$ at locations $s_1, \ldots, s_n$ is
\begin{align}
	\label{eq1}
	F_{HEVP}(c_1,\ldots,c_n;{s_1,\ldots,s_n})&=
	\mathrm{P}\{X_t(s_i)<c_i, i=1, \ldots, n\} \\
	&= \exp\left[-\sum_{l=1}^{L}\left\{\sum_{i=1}^{n}\left(\frac{\omega_l(s_i)}{c_i}\right)^{1/\alpha}\right\}^\alpha\right]. \nonumber
\end{align}
The tail index and the tail dependence are two commonly used measures for the tail behavior of the marginal and the joint distributions.
The tail index is defined as $a=\lim\inf\limits_{c\to\infty}[-\log\{1-{F}(c)\}]/{\log(c)}$.
It is straightforward to show that the HEVP model has tail index $a_{HEVP}=1$.
The tail dependence between $X_t(s_i)$ and $X_t(s_j)$ is
\begin{equation}
\chi_{HEVP}(s_i, s_j) = 2 - \sum_{l=1}^{L}\{\omega_l(s_i)^{1/\alpha}+\omega_l(s_j)^{1/\alpha}\}^{\alpha}\geq 0,
\end{equation}
where $\chi_{HEVP}(s_i,s_j)=\lim_{u\to 1}P\{X_t(s_i)>F_{HEVP}^{-1}(u;s_i)|X_t(s_j)>F_{HEVP}^{-1}(u;$ $s_j)\}$. %
Since $\chi_{HEVP}(s_i, s_j)>0$ assuming both $\omega_l(s_i)$ and $\omega_l(s_j)$ are positive for at least one $l$, $l=1,\ldots, L$,
this model is asymptotically dependent.

\subsection{A semi-parametric Bayesian model for spatial extremes}\label{s2:sb}
On the way towards constructing the HEVP model, the random effects were assigned a parametric distribution $A_{lt}\stackrel{i.i.d.}{\sim} G$, where $G$ is the positive stable distribution.
To allow for more flexibility, the random effects can be modeled semi-parametrically. 
Generalizing further, instead of modeling $A_{1t}, \ldots, A_{Lt}$ independently, we can consider a joint model for $\mathbf{A}_t = (A_{1t},\ldots,A_{Lt})$. 
We model $G$ as a discrete mixture distribution with $J$ terms,
\begin{align*}
	\mathbf{A}_t|\bm \gamma_1,\ldots,\bm \gamma_J &\stackrel{i.i.d.}{\sim} \sum_{j=1}^{J}\pi_j\delta(\bm \gamma_j)\nonumber\\
	\gamma_{jl} &\stackrel{i.i.d.}{\sim} \mathrm{PS}(\alpha)\nonumber
\end{align*}
where $\bm \gamma_j = (\gamma_{1j},\ldots,\gamma_{Lj}), \pi_j=v_i\prod_{j=1}^{i-1}(1-v_j), v_i\stackrel{i.i.d}{\sim} Beta(1,\nu)$ and $\delta({\bm \gamma_j})$ is the Dirac distribution with point mass at ${\bm \gamma_j}$. 
If $J$ is infinite, then this is the Stick-Breaking (SB) representation \citep{sethuraman1994constructive} of the non-parametric Dirichlet process prior \citep{ferguson1974prior} for $G$. 
The prior for $G$ spans the full class of density functions with positive support.
In practice we take $J$ to be finite by setting $v_J=1$ giving a semi-parametric truncated Dirichlet process prior.

Denoting $\bm \gamma = \{\bm \gamma_1,\ldots,\bm \gamma_J\}$, the marginal distribution of the residual process $X_t(\cdot)$ at site $s$ is 
\begin{align}
	\label{eq7}
	F_{SB}(c;s|\bm \gamma)=\mathrm{P}\{X_t(s)<c|\bm \gamma\} = \sum_{j=1}^{J}\pi_j\exp\left[-\left\{\sum_{l=1}^{L}\left(
	\frac{\omega_l(s)}{c}\right)^{1/\alpha} \gamma_{lj}\right\}\right]
\end{align}
and the joint distribution of $X_t(s_1),\ldots, X_t(s_n)$ is
\begin{align}
	\label{eq8}
	F_{SB}(c_1,\ldots,c_n;{s_1,\ldots,s_n}|\bm \gamma)&=
	\mathrm{P}\{X_t(s_i)<c_i, i=1, \ldots, n | \bm \gamma\}\\
	&= 
	\sum_{j=1}^J\pi_j\exp\left[-\sum_{l=1}^{L}
	\left\{\sum_{i=1}^n \left(
	\frac{\omega_l(s_i)}{c_i}
	\right)^{1/\alpha}
	\right\}\gamma_{lj}
	\right].\nonumber
\end{align}

Marginalizing over ${\bm \gamma}$, the expected values of $F_{SB}(c,s|\bm \gamma)$ and $F_{SB}(c_1,\ldots,c_n$; ${s_1,\ldots,s_n}|\bm \gamma)$ are
\begin{align}
	E[F_{SB}(c;s|\bm \gamma)] &= F_{HEVP}(c;s)\\
	E[F_{SB}(c_1,\ldots,c_n;{s_1,\ldots,s_n}|\bm \gamma)] &= F_{HEVP}(c_1,\ldots,c_n;{s_1,\ldots,s_n}).\nonumber
\end{align}
Therefore, this semiparametric model is stochastically-centered on the HEVP model.

Appendix A.2.1 shows that the tail index of the marginal distribution of $X_t(s)$ is $a_{SB}=1/\alpha$ and the tail dependence is $\chi_{SB}(s_i,s_j)=0$ where $\chi_{SB}(s_i,s_j)=\lim_{u\to 1}P\{X_t(s_i)>F_{SB}^{-1}(u;s_i|\bm \gamma)|X_t(s_j)>F_{SB}^{-1}(u;s_j|\bm \gamma),\bm \gamma\}$. Since $\alpha\in(0,1)$, the SB model has a thinner tail than the HEVP model. Unlike the HEVP model that holds a positive tail dependence, the SB model is more appropriate when the tail dependence vanishes. Further, 
max-stability does not hold for the SB model.

\subsection{Max-Mixture hybrid model}\label{s3:mm}
In Section \ref{s1:review} and \ref{s2:sb} we introduced asymptotically dependent and independent models. 
Therefore, if the dependence structure is known in advance, the HEVP or SB model can be chosen accordingly. However, it is generally hard to determine the asymptotic dependence structure in advance. To bridge these dependence classes and maintain flexibility, we propose a hybrid model that combines the strengths of the HEVP and SB models. In addition, the model achieves a smooth transition between the two dependence classes. The Max-Mixture (MM) hybrid model is
\begin{align}
	\label{MM}
	X_{t}(s) = \max\{q\Tilde{X}_{t}(s)^q,(1-q)\hat{X}_{t}(s)^{1-q}\},
\end{align}
where $\Tilde{X}_{t}(s)$ follows the HEVP model in Section 2.1, $\hat{X}_{t}(s)$ follows the SB model in Section 2.2, and $\Tilde{X}_{t}(s)$ and $\hat{X}_{t}(s)$ are independent. The contribution of each process is controlled by the parameter $q\in[0,1]$, which can be inferred by the data. When $q=1$, the model reduces to the HEVP model and when $q=0$ the model reduces to the SB model. 

Conditional on $\bm \gamma$, the marginal distribution is
\begin{align}
	\label{eq5}
	F_{MM}(c;s|\bm \gamma)&=\mathrm{P}(X_t(s)<c|\bm \gamma) \\
	&= F_{HEVP}\left\{\left(\frac{c}{q}\right)^{\frac{1}{q}};s\right\}F_{SB}\left\{\left(\frac{c}{1-q}\right)^{\frac{1}{1-q}};s \middle\vert\bm\gamma\right\} \nonumber.
\end{align}
When $q>\frac{\alpha}{1+\alpha}$, the tail index is $a_{MM}=\frac{1}{q}$ and when $q<\frac{\alpha}{1+\alpha}$, the tail index is $a_{MM}=\frac{1}{\alpha(1-q)}$. Therefore, $a_{HEVP}\leq a_{MM} \leq a_{SB}$ for $q\geq\alpha$ and $a_{HEVP}\leq a_{SB} \leq a_{MM}$ for $q<\alpha$.         

The joint distribution of $X_t(s_1),\ldots,X_t(s_n)$ is
\begin{align}\label{eq666}
	&F_{MM}(c_1,\ldots,c_n;{s_1,\ldots,s_n}|\bm \gamma)\\
	=&\mathrm{P}(X_t(s_1)<c_1,\ldots, X_t(s_n)<c_n |\bm \gamma) \nonumber\\
	=& F_{HEVP}\left\{\left(\frac{c_1}{q}\right)^{\frac{1}{q}},\ldots,\left(\frac{c_n}{q}\right)^{\frac{1}{q}};{s_1,\ldots,s_n}\right\}\nonumber\\
	&\times F_{SB}\left\{\left(\frac{c_1}{1-q}\right)^{\frac{1}{1-q}},\ldots,\left(\frac{c_n}{1-q}\right)^{\frac{1}{1-q}};{s_1,\ldots,s_n} \middle\vert\bm \gamma\right\} \nonumber
\end{align}
where $F_{HEVP}$ and $F_{SB}$ are given in \eqref{eq1} and \eqref{eq8}.
The tail dependence of the proposed model is
\[\chi_{MM}(s_1,s_2)= \begin{cases} 
0, & q<\frac{\alpha}{1+\alpha} \\
2-\sum_{l=1}^{L}(\omega_l(s_1)^{1/\alpha}+\omega_l(s_2)^{1/\alpha})^{\alpha}, & q>\frac{\alpha}{1+\alpha} 
\end{cases}
\]
where $\chi_{MM}(s_i,s_j)=\lim_{u\to 1}P\{X_t(s_i)>F_{MM}^{-1}(u;s_i|\bm \gamma)|X_t(s_j)>F_{MM}^{-1}(u;$ $s_j|\bm \gamma),\bm \gamma\}$.
The detailed derivation is postponed to Appendix A.2.2. Thus the proposed hybrid model can capture both the asymptotic independence and dependence structures and achieve a smooth transition between the two sub-families. Therefore, 
\[\delta=I\left(q\geq\frac{\alpha}{1+\alpha}\right)\]
is an indicator of asymptotic dependence, and the posterior probability that $\delta=1$ is a measure of evidence for asymptotic dependence.

\section{Simulation study}\label{s:sim}
\subsection{Simulation settings}\label{s:sim:setting}
In this section, we evaluate the numerical performance of the models described in Section \ref{s:method}. We generate 50 data sets from each of the following six settings:
\begin{itemize}
	\item[(1)] Max-stable process (MS)
	\item[(2)] Stick-breaking process (SB)
	\item[(3)] Gaussian process (GP)
	\item[(4)] Skewed-$t$ process (ST)
	\item[(5)] Inverted max-stable process (InvMS)
	\item[(6)] Max Gaussian-MS mixture process (MAX)
\end{itemize}
In each case, data are generated at $n=49$ locations on a $7\times 7$ grid covering $[1,7]\times[1,7]$ and with $T=50$ independent replicates. 
For settings (1) and (2), we generate data from the HEVP model and the SB model with 
kernels $\omega_l(s)\propto \exp\left(-\frac{(s-v_l)^2}{2\tau^2}\right)$ for $l=1,\ldots,L$, where $v_l$ are spatial knots that are set to be the same with the grid and the kernel bandwidth $\tau=1$. GEV parameters $\mu(s)=0.1, \sigma(s)=1, \xi(s)=0.1, \alpha=0.3$. In Setting (2) we let the number of components be $J=3$ and ${\bm \pi}=(0.5,0.3,0.2)$.
For Setting (3), we generate data $Y_t(s)$ from the Gaussian process with mean 0.1, variance 1 and exponential spatial correlation $\mathrm{cor}\{Y_t(s_i),Y_t(s_j)\}=\exp(-\|s_i-s_j\|)$. For Setting (4), we generate data from Skew-$t$ process $Y_t(s)=\mu(s)+\lambda\sigma_t|Z_t|+\sigma_te_t(s)$, where the skewness parameter is $\lambda=3$, $\mu(s)=1$, $Z_t\sim N(0,1)$, $\sigma_t^2 \sim InvGamma(4,1)$ and $e_t(s)$ is a Gaussian process with mean 0, variance 1 and exponential spatial correlation $\exp(-\|s_i-s_j\|)$. The Gaussian process is asymptotically independent for both upper and lower tails, while the skew-t process is asymptotically dependent in both tails, but not max-stable.
For Setting (5), we generate data from the inverted max-stable process \citep{wadsworth2012dependence} with unit exponential marginal distributions, i.e., the residual process $X_t(s)$ is a max-stable process with unit Fr{\'e}chet marginal distributions generated as in Setting (1). The data are then $Y_t(s)=\mu(s)+\frac{\sigma(s)}{\xi(s)}[X_t(s)^{-\xi(s)}-1]$ with $\mu(s)=0.1, \sigma(s)=1$ and $\xi(s)=0.1$. This process is asymptotically dependent in the lower tail but asymptotically independent in the upper tail.
For Setting (6), we generate data from a mixture of MS and Gaussian, i.e., the residual process $X_t(s)=\max\{qX_{1t}(s)^{q}, (1-q)X_{2t}(s)^{(1-q)}\}$, $X_{1t}(s)$ and $X_{2t}(s)$ are generated under settings (1) and (3), respectively. The data are then $Y_t(s)=\mu(s)+\frac{\sigma(s)}{\xi(s)}[X_t(s)^{\xi(s)}-1]$ with $\mu(s)=0.1, \sigma(s)=1$ and $\xi(s)=0.1$. In the simulation, we fix $q=0.5$. The process is tail dependent in the upper tail.

For each data set, we fit the HEVP, SB and MM model.
We use Markov chain Monte Carlo (MCMC) methods for model fitting.
Gaussian process priors are used for the GEV parameters.
We generate 10,000 MCMC iterations and discard the first 2,500 as burn-in. For the SB and MM model, we fix the number of components in the stick-breaking prior be the same with the number of replicates, $J=50$, by setting $v_J=1$. Prior distributions and computational details are given in Appendix A.1.

We compare models using the estimated marginal quantiles and estimated pairwise tail dependence. We fit the models to $n=49$ sites and compute the true quantiles $Q_{\kappa}(s_i)$ for site $s_i$ at quantile level $\kappa$ and the true pairwise tail dependence $\hat{\chi}_{u}(s_i,s_j)$ for site $s_i$ and $s_j$ at quantile level $u$ and $i\in\{1,2,.\ldots,n\}$. The MSE of $\hat{Q}_{\kappa,b}(s_i)$ and $\hat{\chi}_{u,b}(s_i,s_j)$ for data set $b$ is defined as $\frac{1}{n}\sum_{i=1}^{n}(\hat{Q}_{\kappa,b}(s_i)-Q_{\kappa}(s_i))^2$ and $\frac{1}{m}\sum_{i<j}(\hat{\chi}_{u,b}(s_i,s_j)-\chi_{u,b}(s_i,s_j))^2$, where $m=n(n-1)/2$ is the number of pairs of sites. Figures 1 and 2 plot the mean MSE (MMSE) over 50 data sets, i.e.,
\begin{align}
\label{value}
MMSE(\hat{Q}_{\kappa}) &= \frac{1}{50}\sum_{b=1}^{50}\left\{\frac{1}{n}\sum_{i=1}^{n}(\hat{Q}_{\kappa,b}(s_i)-Q_{\kappa}(s_i))^2\right\} \\
MMSE(\hat{\chi}_u) &= \frac{1}{50}\sum_{b=1}^{50}\left\{\frac{1}{m}\sum_{i<j}(\hat{\chi}_{u,b}(s_i,s_j)-\chi_{u,b}(s_i,s_j))^2\right\}. \nonumber
\end{align}
Appendix A.3 reports these values in a table with standard errors.

\subsection{Simulation results}\label{s:sim:result}
\begin{figure}
	\centering
	\includegraphics[width=0.96\linewidth]{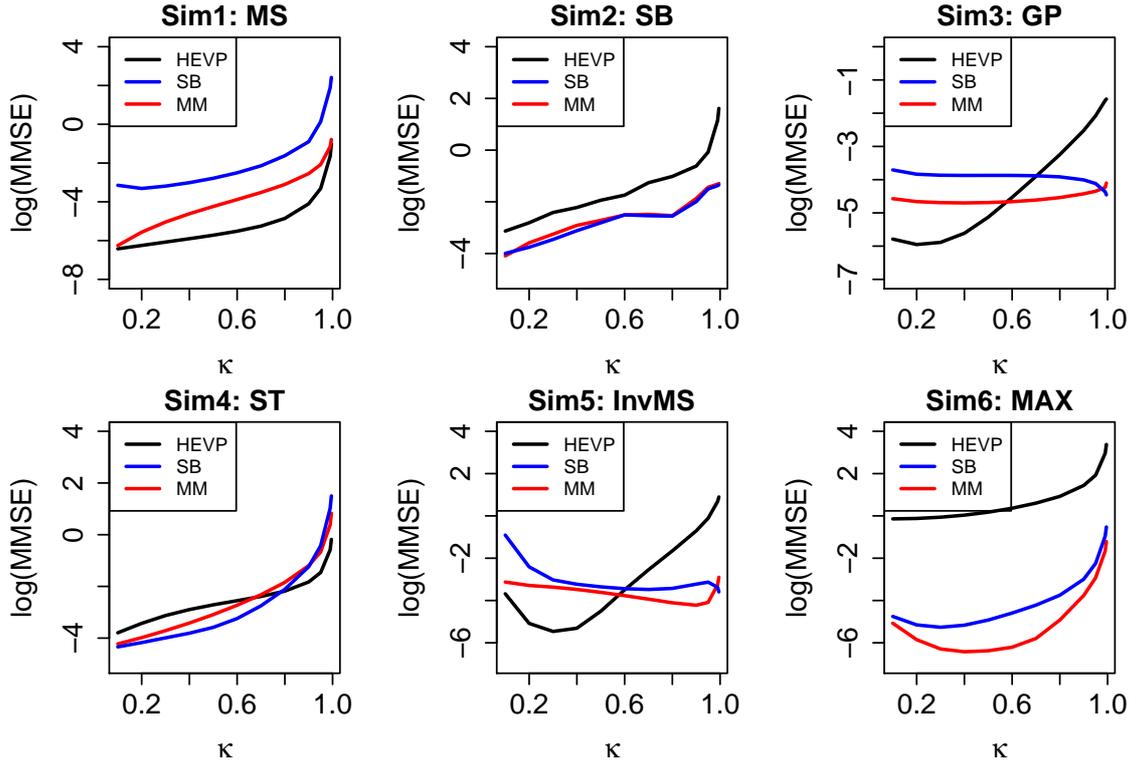}
	\caption{$\log\left[\mathrm{MMSE}(\hat{Q}_{\kappa})\right]$ for quantile levels $\kappa\in \{0.1,0.2,\ldots, 0.8,0.9, 0.95, 0.99$, $0.995\}$. ``HEVP" stands for the hierachical extreme value process model, ``SB" stands for the extended stick-breaking prior model, ``MM" stands for the max-mixture hybrid model.}
	\label{fig:qs}
\end{figure}

Figure \ref{fig:qs} plots the log of MMSE$(\hat{Q}_{\kappa})$ by $\kappa$ for the six settings. The HEVP, SB and MM models achieve the smallest MMSE in Settings (1), (2) and (6), respectively, as expected 
since in these settings these models are correctly specified. In Settings (1) and (2), the MM model remains competitive with the top performing models. When the model is misspcified, the HEVP model and the SB model have different patterns in the lower and upper quantiles: In Settings (3) and (5) where the data are tail independent, the SB model delivers better quantile estimates in the tail as expected, however, the HEVP model outperforms the other models in the bulk. In Setting (4) where the data are tail dependent, the pattern is opposite to Settings (3) and (5). 
Our proposed MM model generally has better performance in the bulk for the distribution. In the tail, the MM model has the best performance in Settings (3) and (5), where the model is tail independent, and competitive MMSE in Setting (4), where the data are tail dependent. 
Overall, our proposed MM model delivers the most robust estimation across quantile levels and simulation scenarios.

\begin{figure}
	\centering
	\includegraphics[width=0.96\linewidth]{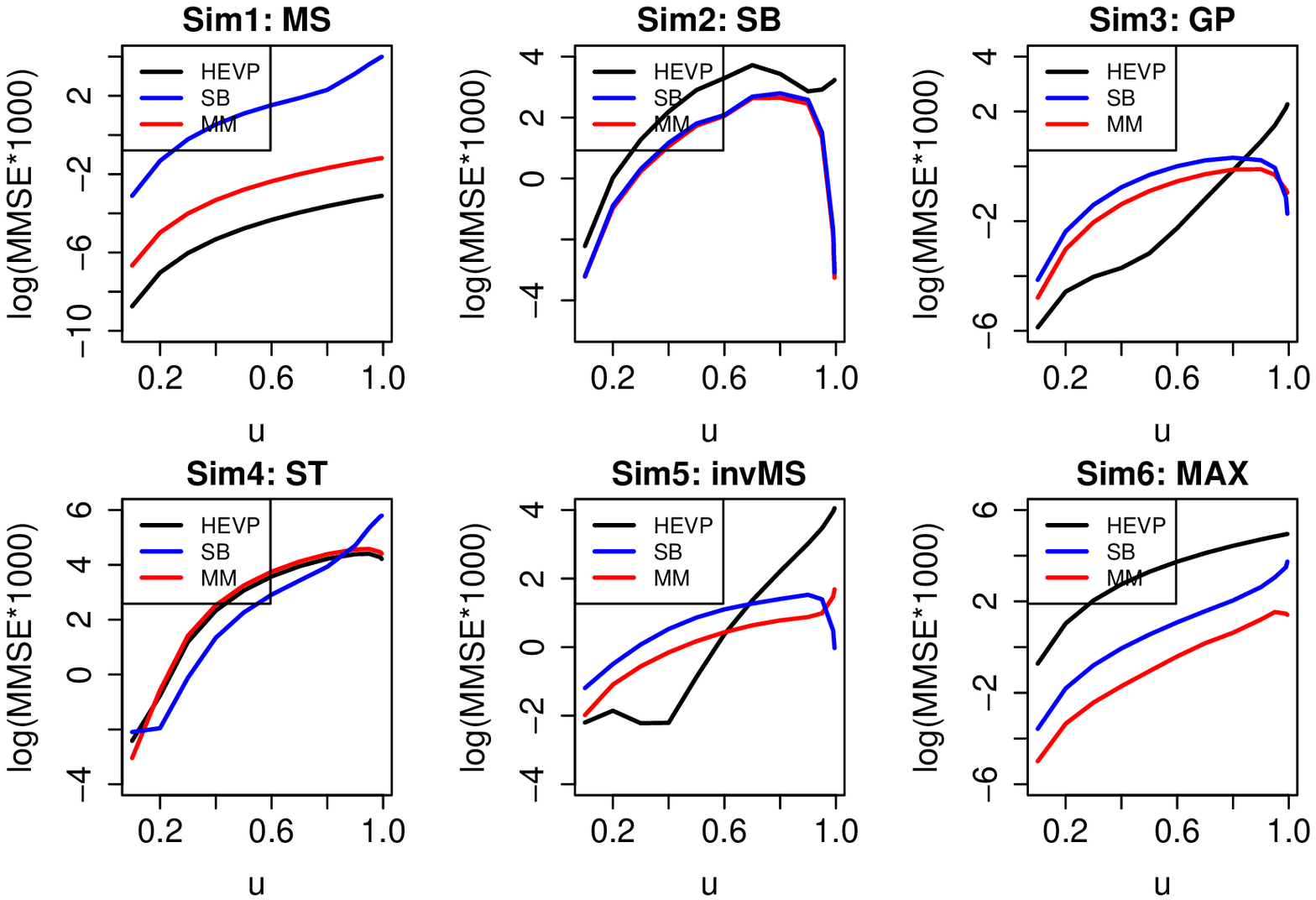}
	\caption{$\log\left[\mathrm{MMSE}(\hat{\chi}_u)\times 1000\right]$ for quantile levels $u\in \{0.1,0.2,\ldots,0.8, 0.9, 0.95,0.99$, $0.995\}$. ``HEVP" stands for the hierachical extreme value process model, ``SB" stands for the extended stick-breaking prior model, ``MM" stands for the max-mixture hybrid model.}
	\label{fig:chi}
\end{figure}

Figure \ref{fig:chi} plots the log of MMSE$(\hat{\chi}_u)\times 1000$ by $u$ for six settings. 
As for $\hat{Q}_\kappa$, the correctly specified models give the best performance in Settings (1), (2) and (6).
In Settings (3) and (5) where the data are tail independent, the HEVP model tends to over estimate the tail dependence when $u$ is close to 1 and thus has the highest MMSE. 
Therefore, similar to the results for marginal quantiles, our proposed MM model delivers most robust estimation over the full range of pairwise tail dependence.

\begin{table}[ht]
	\centering
	\caption{Mean and standard deviation of the posterior probability of  asymptotic dependence (i.e. $\delta=1$) in the MM model over 50 simulation data sets.}
	\label{tab:sim}
	\begin{tabular}{ccccccc}
		\toprule
		Setting & MS & SB & GP & ST & InvMS & MAX \\ 
		\midrule
		Mean & 1.000 & 0.002 & 0.053 & 1.000 & 0.515 & 1.000 \\ 
		SD & 0.000 & 0.021 & 0.184 & 0.000 & 0.459 & 0.000\\ 
		\bottomrule
	\end{tabular}
\end{table}

As we discussed previously, $\delta=I\{q\geq\alpha/(1+\alpha)\}$ indicates tail dependence. In our Bayesian analysis we obtain the posterior probability of tail dependence as the posterior probability that $\delta=1$. We present the mean and standard deviation of this probability over datasets in Table \ref{tab:sim}. In all settings except Setting (5), the MM model successfully identifies the tail properties with a high confidence level. In Setting (5), the MM model infers the model as tail independent with around probability 0.5. This is a challenging case with dependence in the lower but not upper tail and the results for our model reflect this complexity.

\section{Application to Netherlands Wind Gusts}\label{s:app}
\begin{figure}
	\centering
	\includegraphics[width=0.7\linewidth]{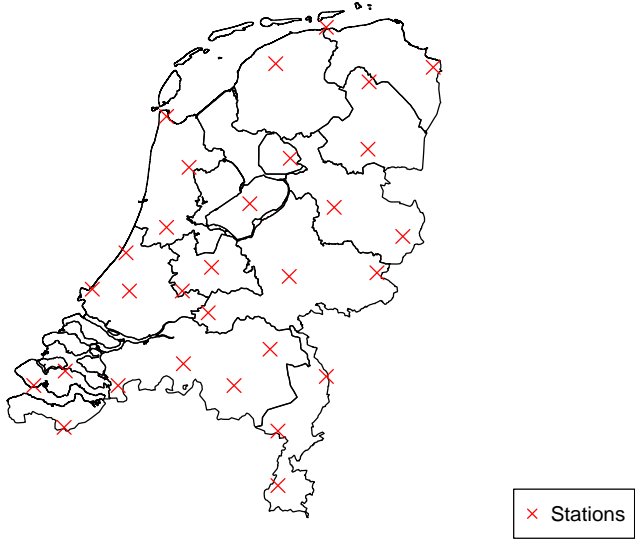}
	\caption{{Climate stations for the Netherlands wind gust data.}}
	\label{fig:data2}
\end{figure}

We applied the proposed models to daily maximum wind gusts data in the Netherlands from 11/14/1999 to 11/13/2008 for 30 climate stations as shown in Figure \ref{fig:data2}. The data are downloaded from the KNMI Climate Explorer website \url{http://climexp.knmi.nl} and were analyzed in \cite{opitz2016modeling}. 
As a renewable energy, wind gust is a key to designing a wind energy conversion system \citep{zervos2006pure, wiser2015wind, scheuerer2015probabilistic}. Other application fields include severe weather forecasting \citep{powers2007numerical,friederichs2012forecast}, air transportation operation \citep{perry1994wind,koetse2009impact}, wind farming \citep{steinkohl2013extreme}, etc. Recent studies have applied both asymptotically dependent \citep{einmahl2016m, oesting2017statistical} and independent models \citep{opitz2016modeling} to the data. 

We assume that the process is stationary in time and with spatially-varying GEV parameters $\mu(s), \sigma(s)$ and $\xi(s)$. We use Gaussian process prior for these parameters. The Gaussian process $\mu({ s})$ has mean $x({s})^T\beta_\mu$ and Mat{\'e}rn covariance where $x({s})=(1,{s})$ indicates latitude and longitude. 
The other GEV paramters $\log\{\sigma(s)\}$ and $\xi(s)$ are modeled similarly.
The knots are set to be the same with the station locations. We conduct a 3-fold cross validation, that is, we randomly split the 30 climate stations into three equal size groups. Each time we fit the model on two groups and evaluate the model on the remaining group. We compare the estimated quantiles and tail dependent coefficient to the empirical values, respectively. Define the mean MSE of estimated quantiles and tail dependent coefficient as
\begin{align*}
MMSE(\hat{Q}_{\kappa}) &= \frac{1}{3}\sum_{b=1}^{3}\left\{\frac{1}{n_b}\sum_{i=1}^{n_b}(\hat{Q}_{\kappa,b}(s_i)-Q^{E}_{\kappa,b}(s_i))^2\right\} \\
MMSE(\hat{\chi}_u) &= \frac{1}{3}\sum_{b=1}^{3}\left\{\frac{1}{M_b}\sum_{i<j}^{{M_b}}(\hat{\chi}_{u,b}(s_i,s_j)-\chi_{u,b}^{E}(s_i,s_j))^2\right\},
\end{align*}
where $Q^{E}_{\kappa,b}(s_i)$ is the empirical $\kappa$-th quantile at site $s_i$ in $b$-th testing set, $\chi_{u,b}^{E}(s_i,s_j)$ is the empirical $u$-th tail dependence between site $s_i$ and site $s_j$ in $b$-th testing set, $M^{(b)}$ is the number of pairs of sites in $b$-th testing set.

\begin{table}[ht]
	\centering
	\caption{Mean MSE of estimated quantiles and tail dependent coefficient. ``HEVP" stands for the hierachical extreme value process model, ``SB" stands for the extended stick-breaking prior model, ``MM" stands for the max-mixture hybrid model.}
	\label{tab:data2}
	\begin{tabular}{c|cccccc}
		\toprule
		& 0.5 & 0.8 & 0.9 & 0.95 & 0.99 & 0.995 \\
		\cline{2-7}
		& \multicolumn{6}{|c}{$MMSE(\hat{Q}_{\kappa})$}\\
		\hline
		HEVP & 0.601 & 0.958 & 1.278 & 2.246 & 18.492 & 32.341 \\ 
		SB & 0.498 & 0.927 & 1.208 & 1.620 & 3.230 & 3.010 \\  
		MM &  0.532 & 0.972  & 1.367  & 1.846  & 4.021  & 4.307 \\
		\midrule
		& \multicolumn{6}{|c}{$MMSE(\hat{\chi}_u)$}\\
		\cline{2-7}
		HEVP & 0.049 & 0.083 & 0.105 & 0.150 & 0.177 & 0.281 \\  
		SB & 0.056 & 0.071 & 0.067 & 0.082 & 0.036 & 0.020 \\  
		MM & 0.055  & 0.071 & 0.068  & 0.077 & 0.015 &  0.016\\ 
		\bottomrule
	\end{tabular}
\end{table}

The SB model has the best MMSE for all estimated quantiles (Table \ref{tab:data2}). The HEVP model has competitive results for lower quantile levels ($\kappa\leq 0.9$). However, for the upper tail quantiles, the HEVP model delivers unreasonable results. The MM model has similar results to the best model across all quantile levels.

The MM model outperforms the others in estimating the tail dependent coefficient except for quantile level $u=0.5$. 
The posterior probability that $q\geq\alpha/(1+\alpha)$ is 0, which strongly suggests asymptotic independence. This agrees with previous studies that claim it is more appropriate to use an asymptotically independent models to fit wind gust data \citep{ledford1996statistics, opitz2013extremes, opitz2016modeling}. Our results verified this proposition and presented two appropriate models (SB and MM) to study the wind gust data. 

\begin{figure}
	\centering  
	\subfloat[0.5 quantile, MS]{\includegraphics[width=0.32\linewidth]{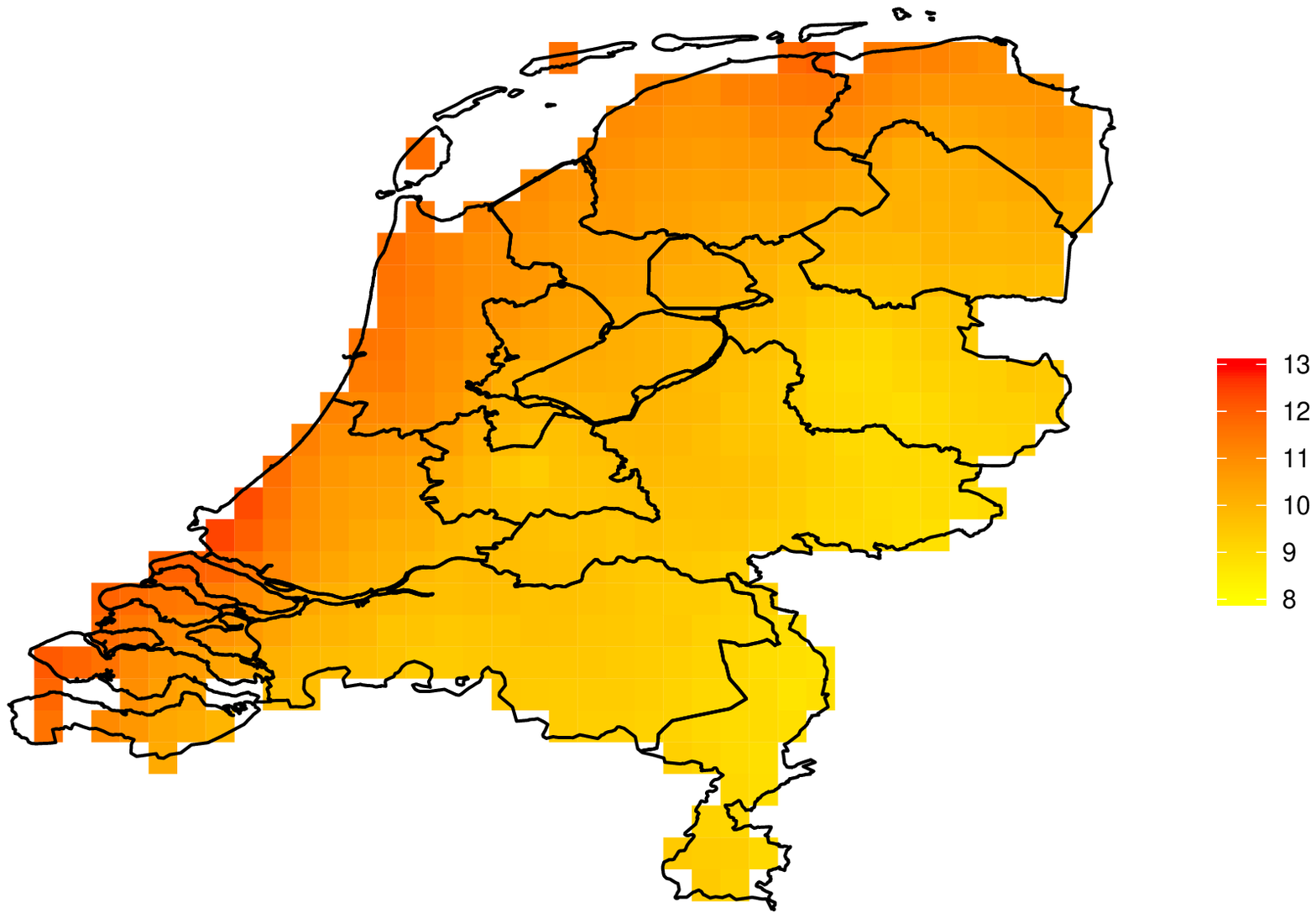}}
	\subfloat[0.5 quantile, SB]{\includegraphics[width=0.32\linewidth]{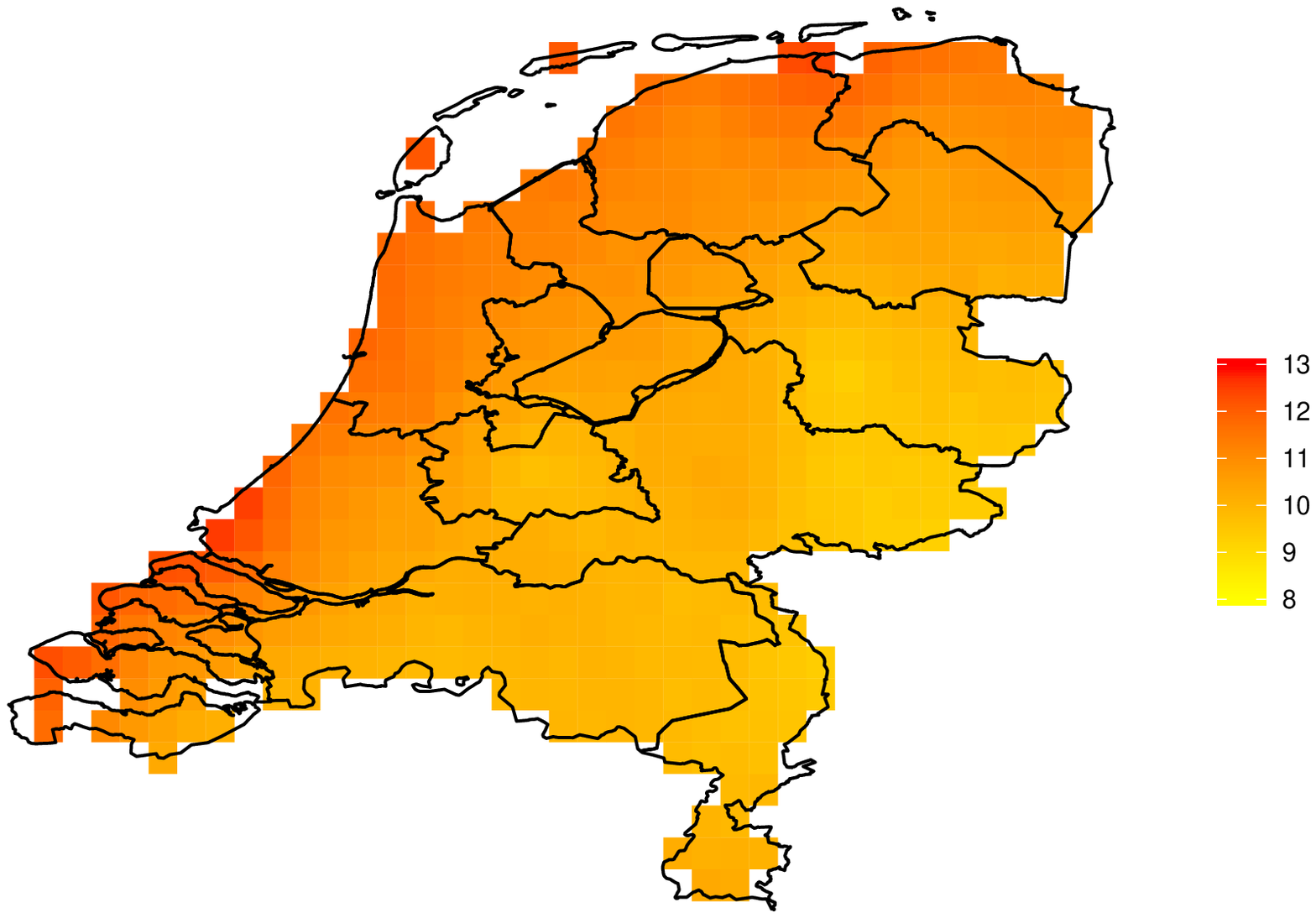}}
	\subfloat[0.5 quantile, MM]{\includegraphics[width=0.32\linewidth]{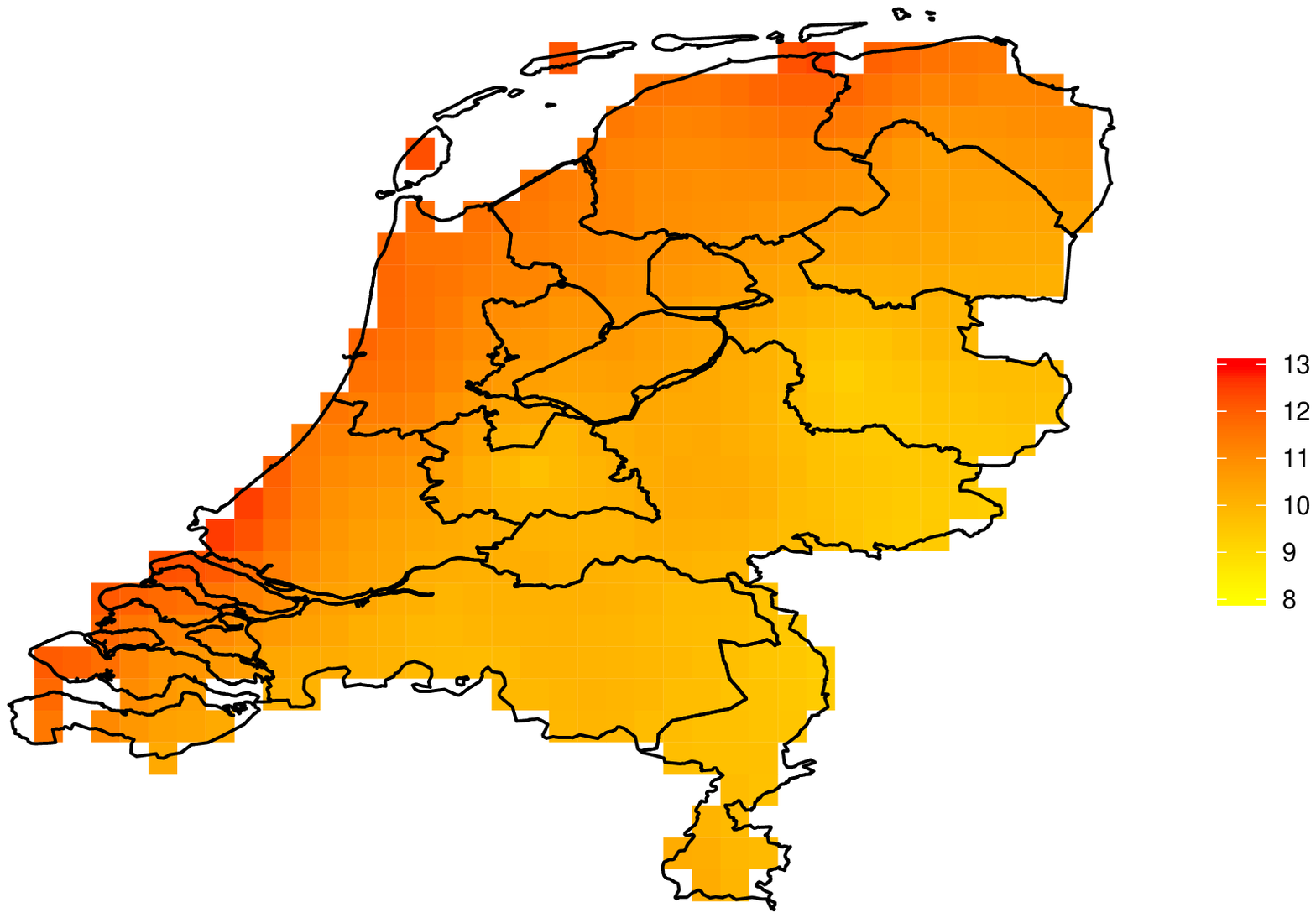}}
	\qquad
	\subfloat[0.95 quantile, MS]{\includegraphics[width=0.32\linewidth]{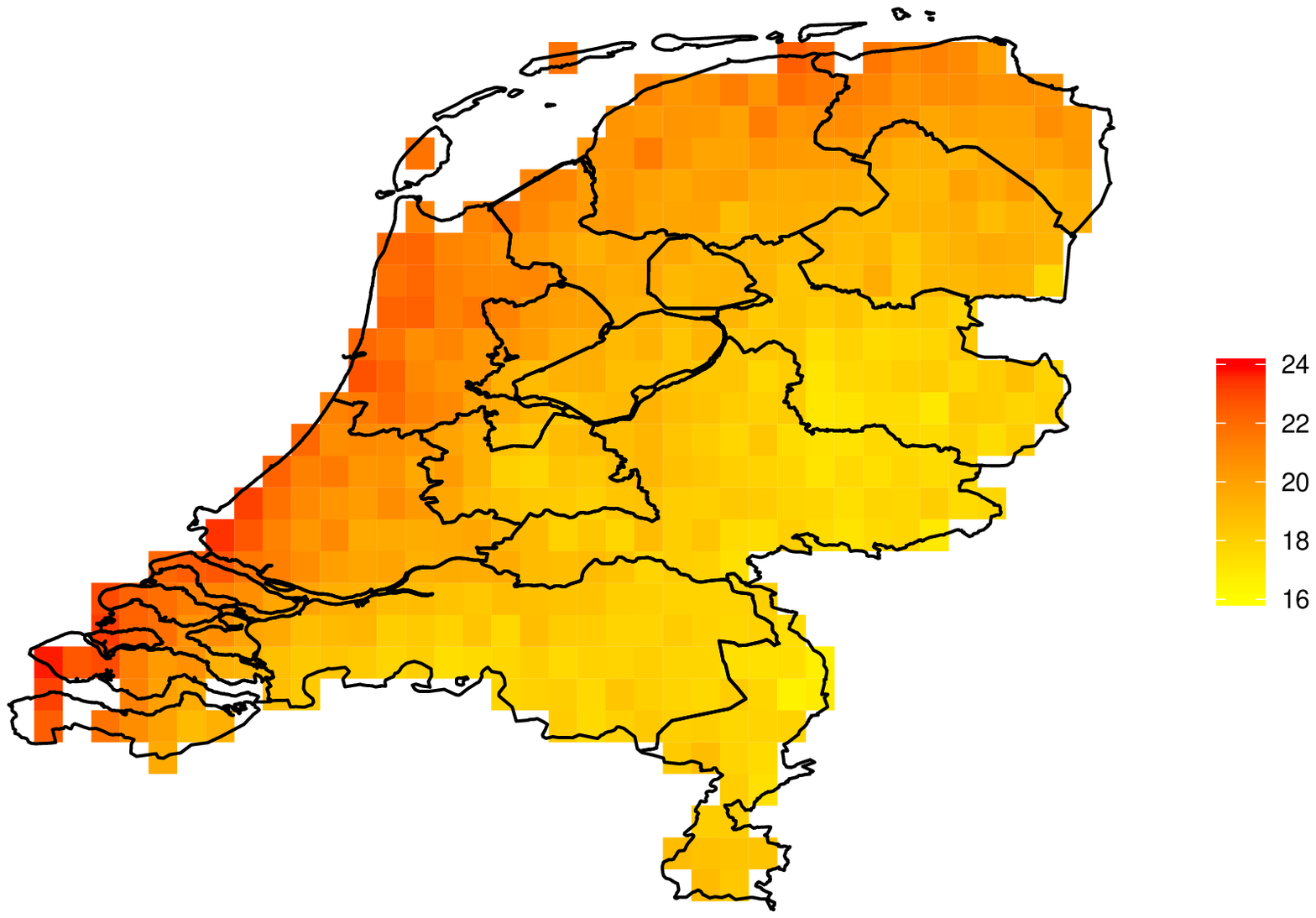}}
	\subfloat[0.95 quantile, SB]{\includegraphics[width=0.32\linewidth]{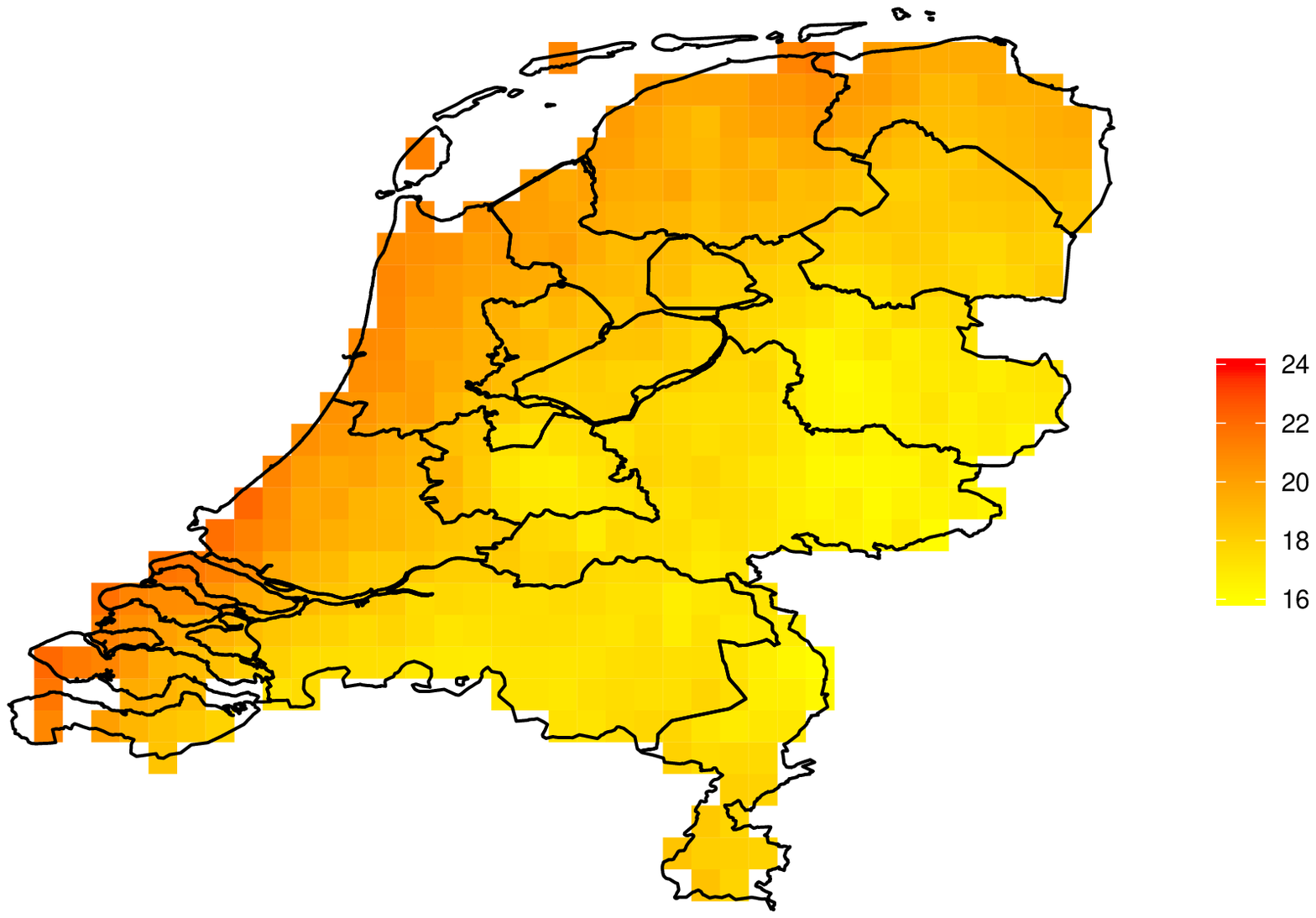}}
	\subfloat[0.95 quantile, MM]{\includegraphics[width=0.32\linewidth]{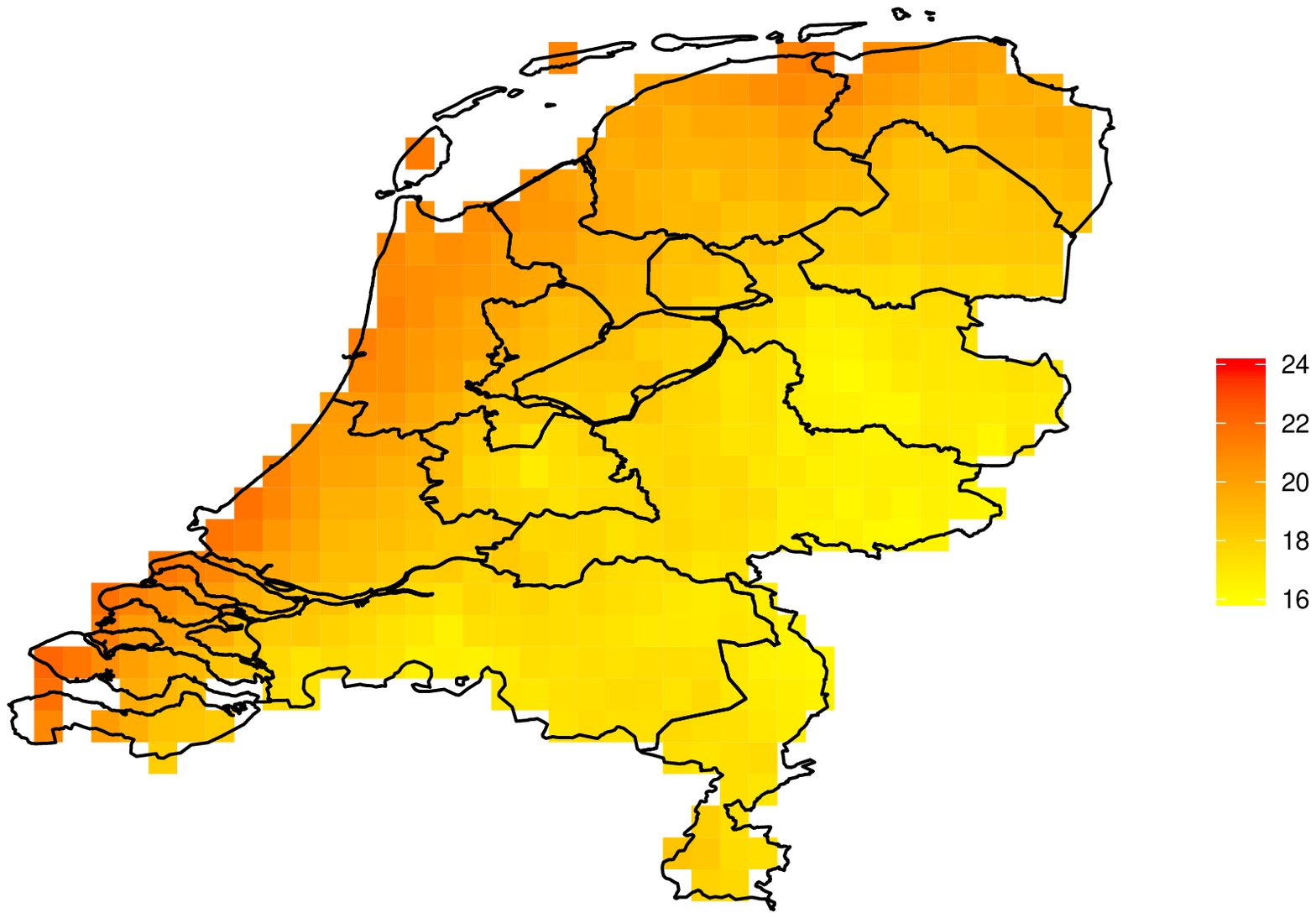}}
	\qquad
	\subfloat[0.99 quantile, MS]{\includegraphics[width=0.32\linewidth]{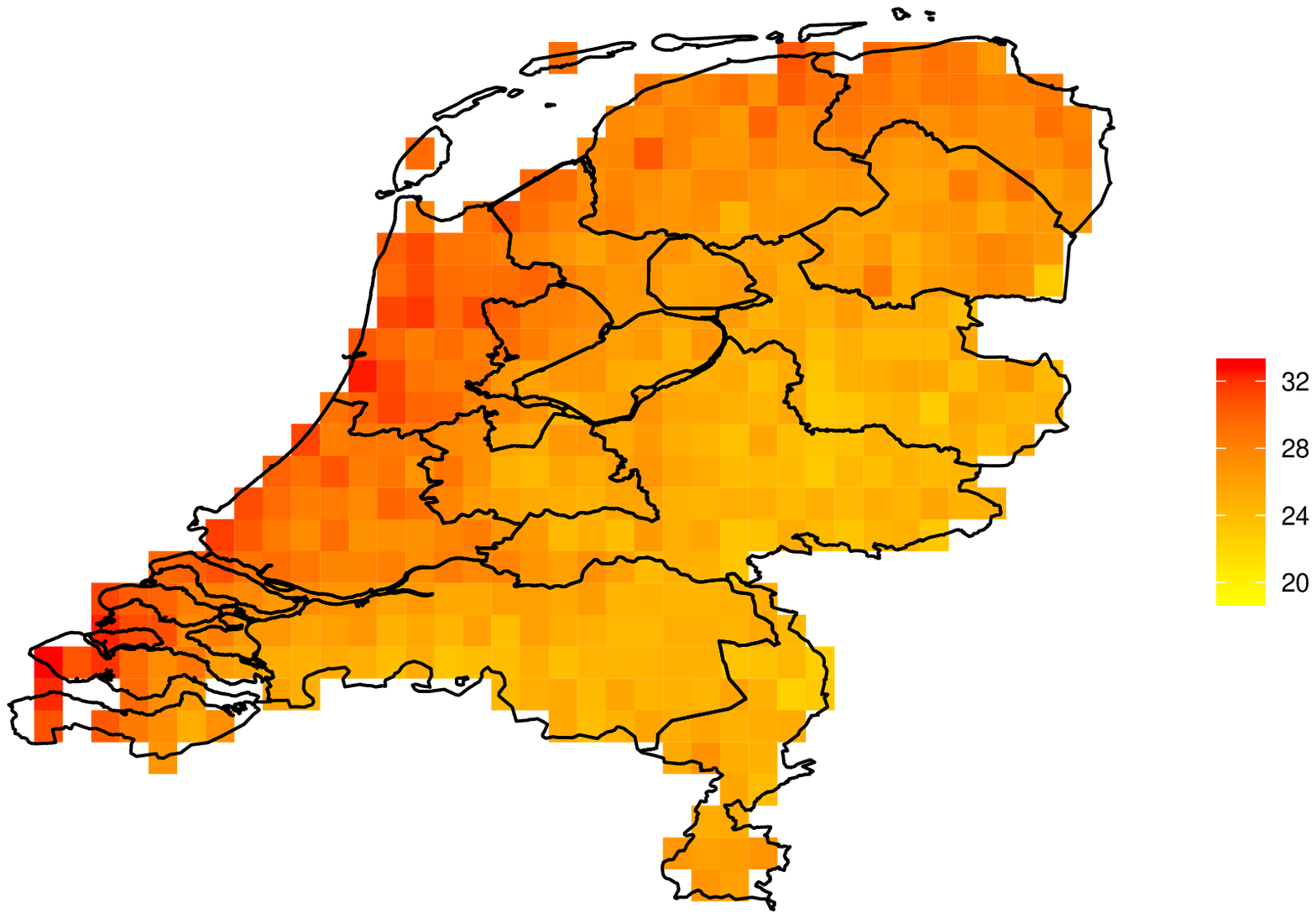}}
	\subfloat[0.99 quantile, SB]{\includegraphics[width=0.32\linewidth]{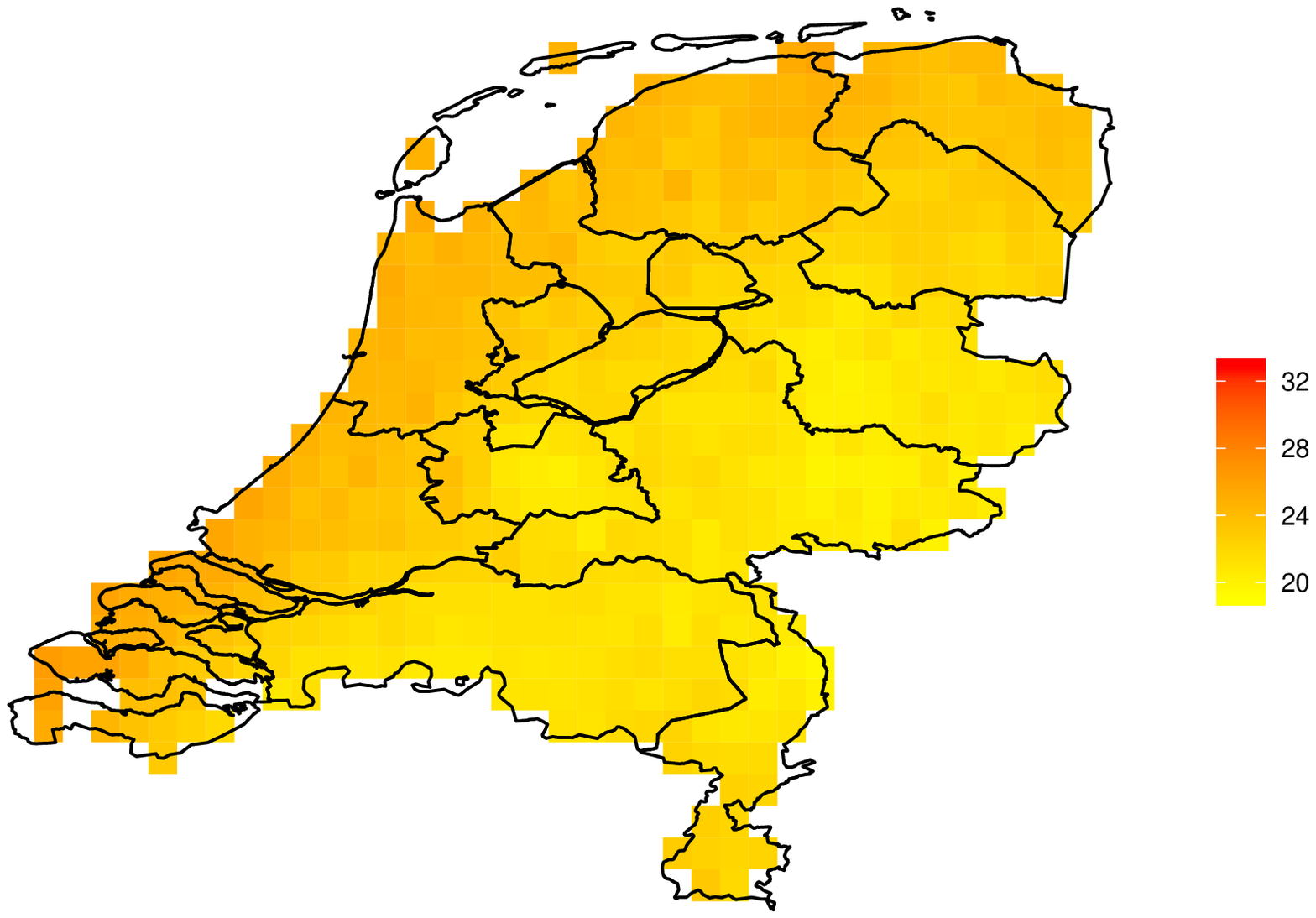}}
	\subfloat[0.99 quantile, MM]{\includegraphics[width=0.32\linewidth]{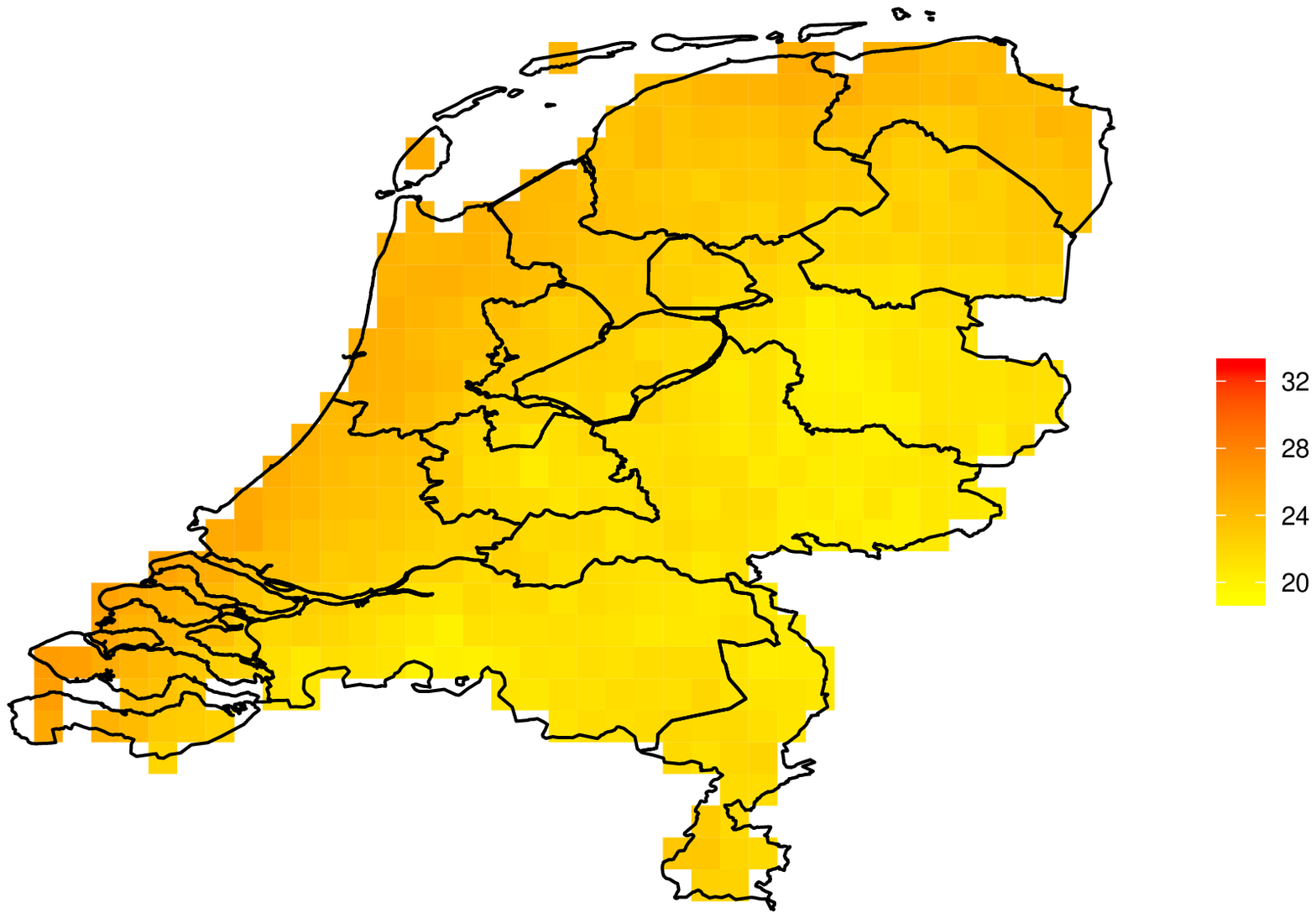}}
	\caption{Posterior mean of the 0.50, 0.95 and 0.99 quantiles for the predicted wind speed. All units are m/s.}
	\label{qmn}
\end{figure}

\begin{figure}
\centering  
\subfloat[0.5 quantile, MS]{\includegraphics[width=0.32\linewidth]{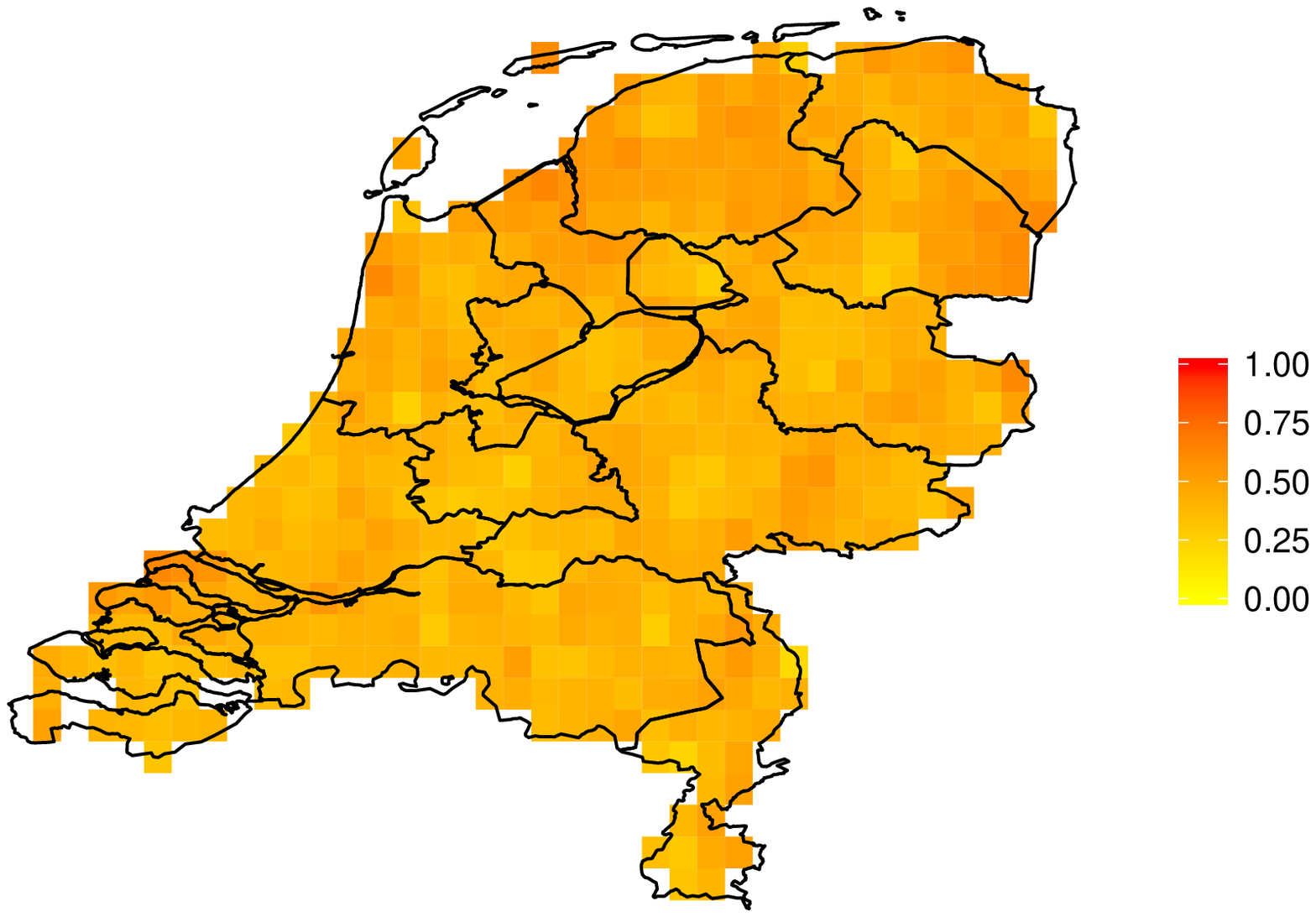}}
\subfloat[0.5 quantile, SB]{\includegraphics[width=0.32\linewidth]{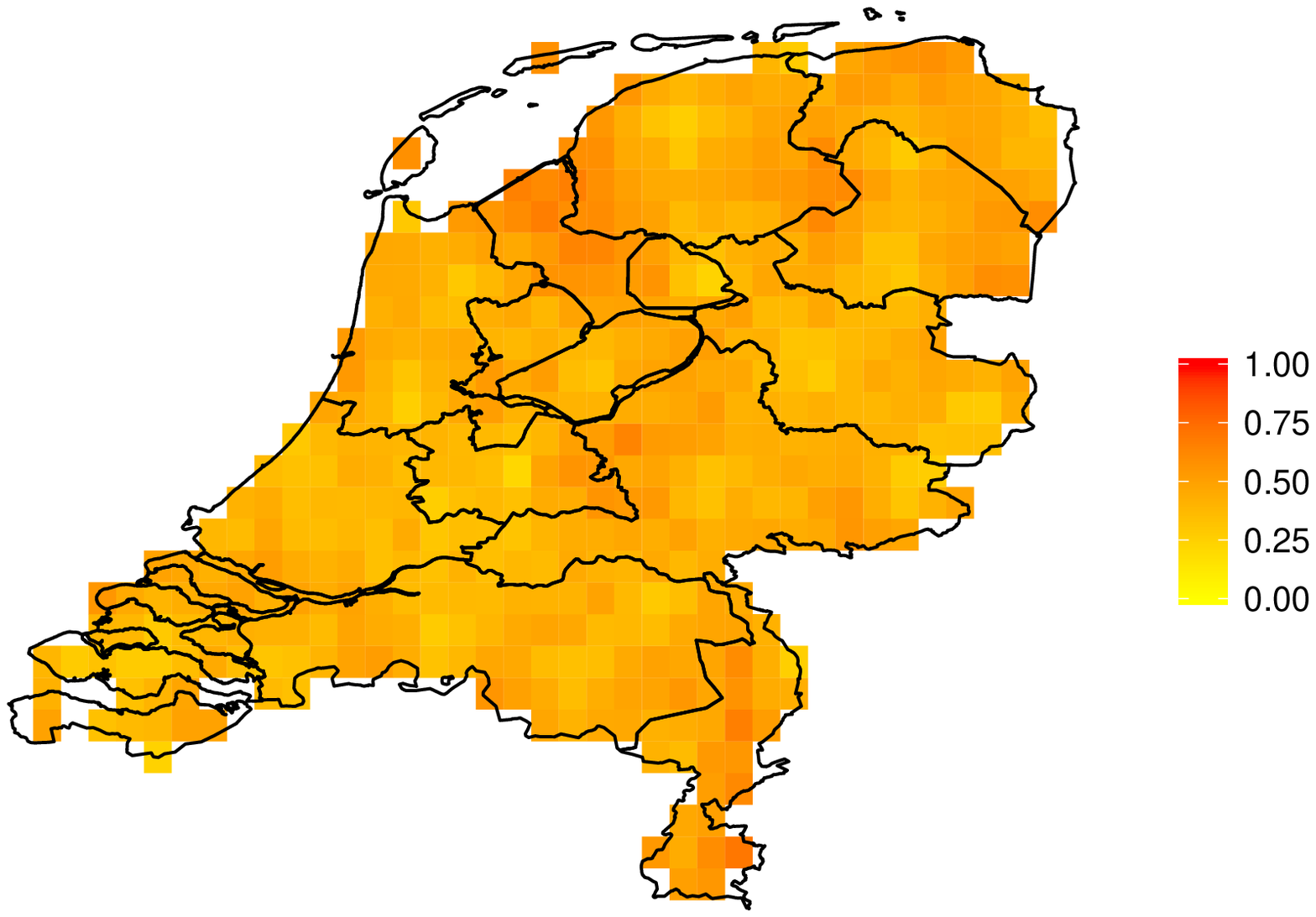}}
\subfloat[0.5 quantile, MM]{\includegraphics[width=0.32\linewidth]{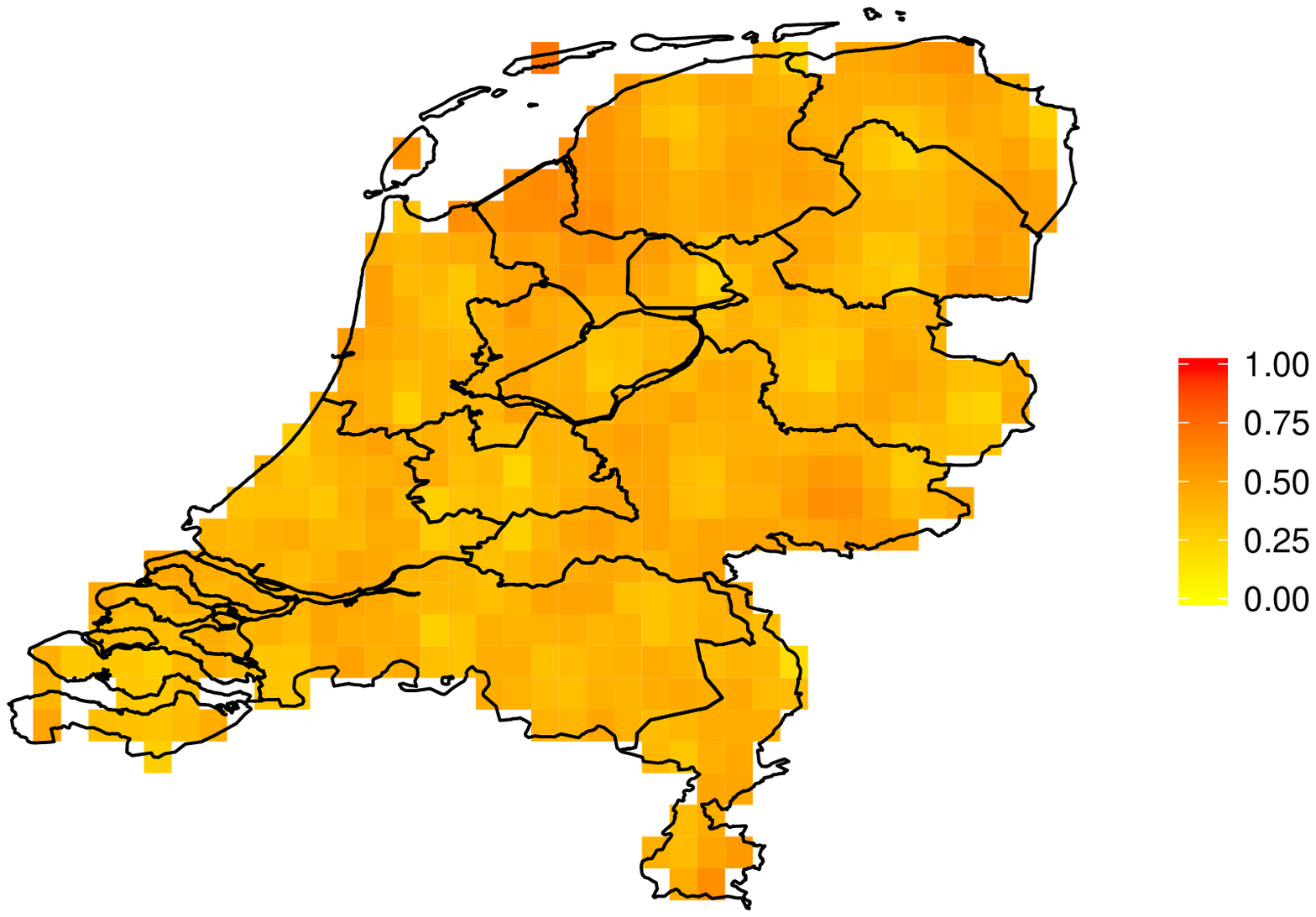}}
\qquad
\subfloat[0.95 quantile, MS]{\includegraphics[width=0.32\linewidth]{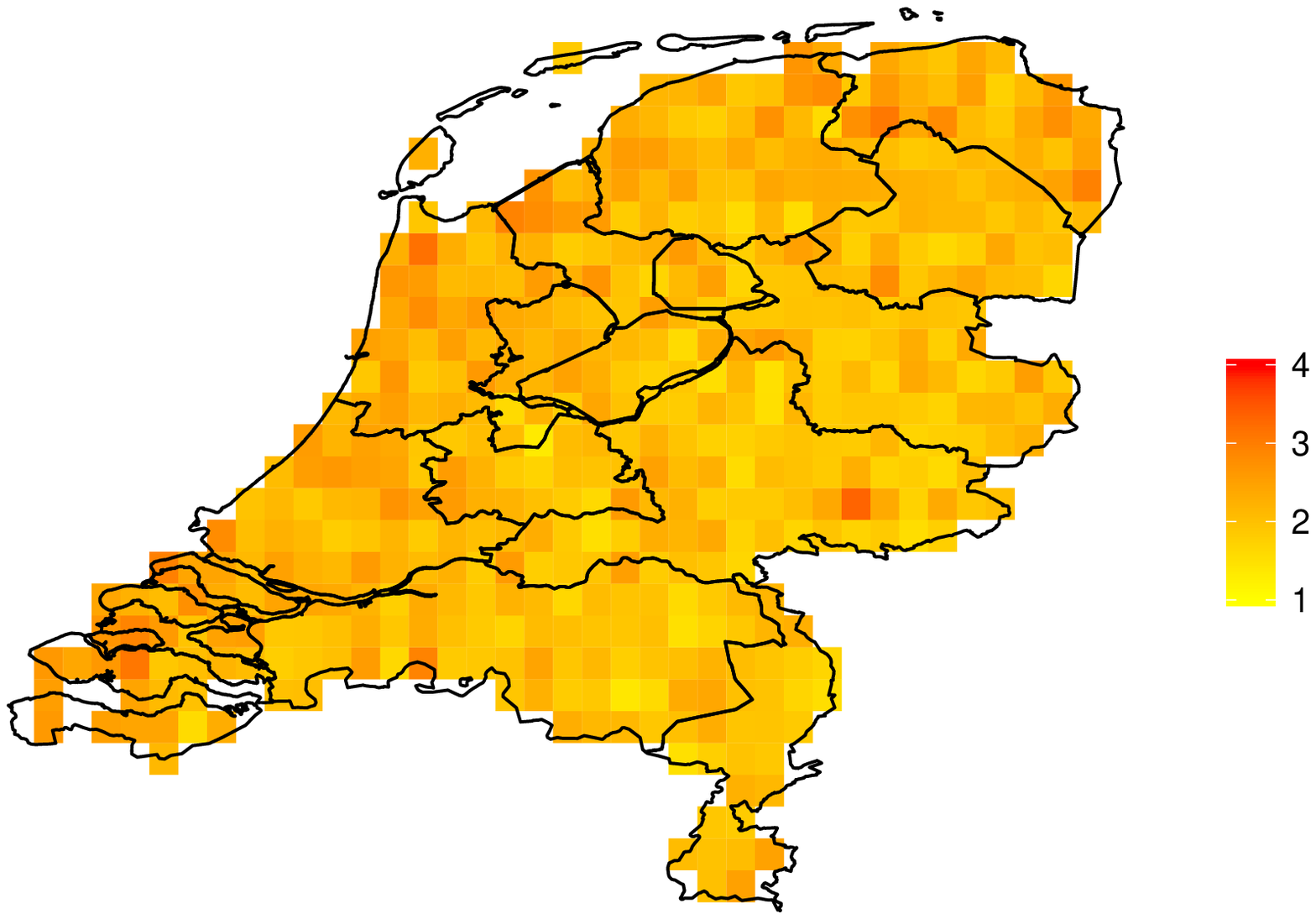}}
\subfloat[0.95 quantile, SB]{\includegraphics[width=0.32\linewidth]{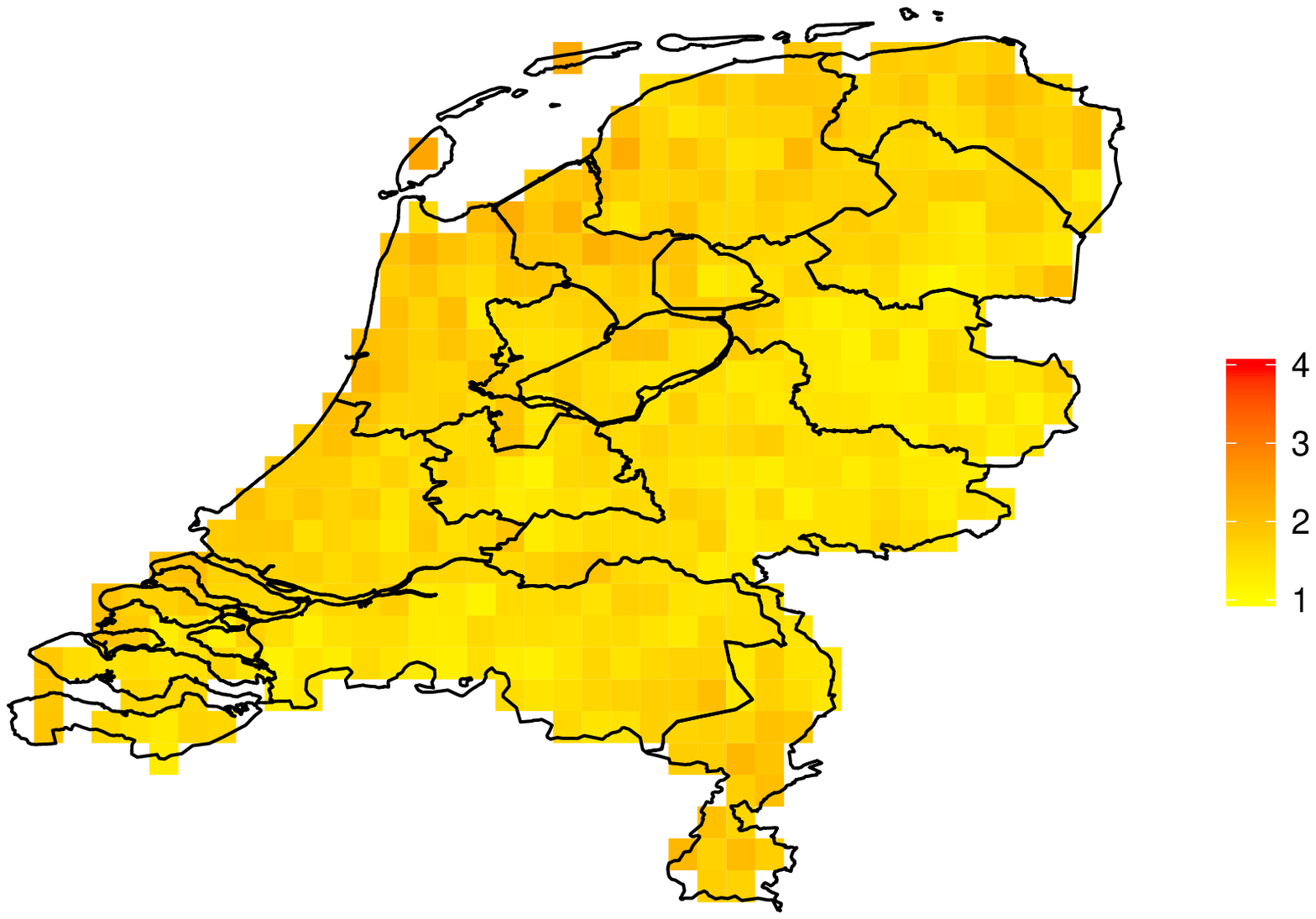}}
\subfloat[0.95 quantile, MM]{\includegraphics[width=0.32\linewidth]{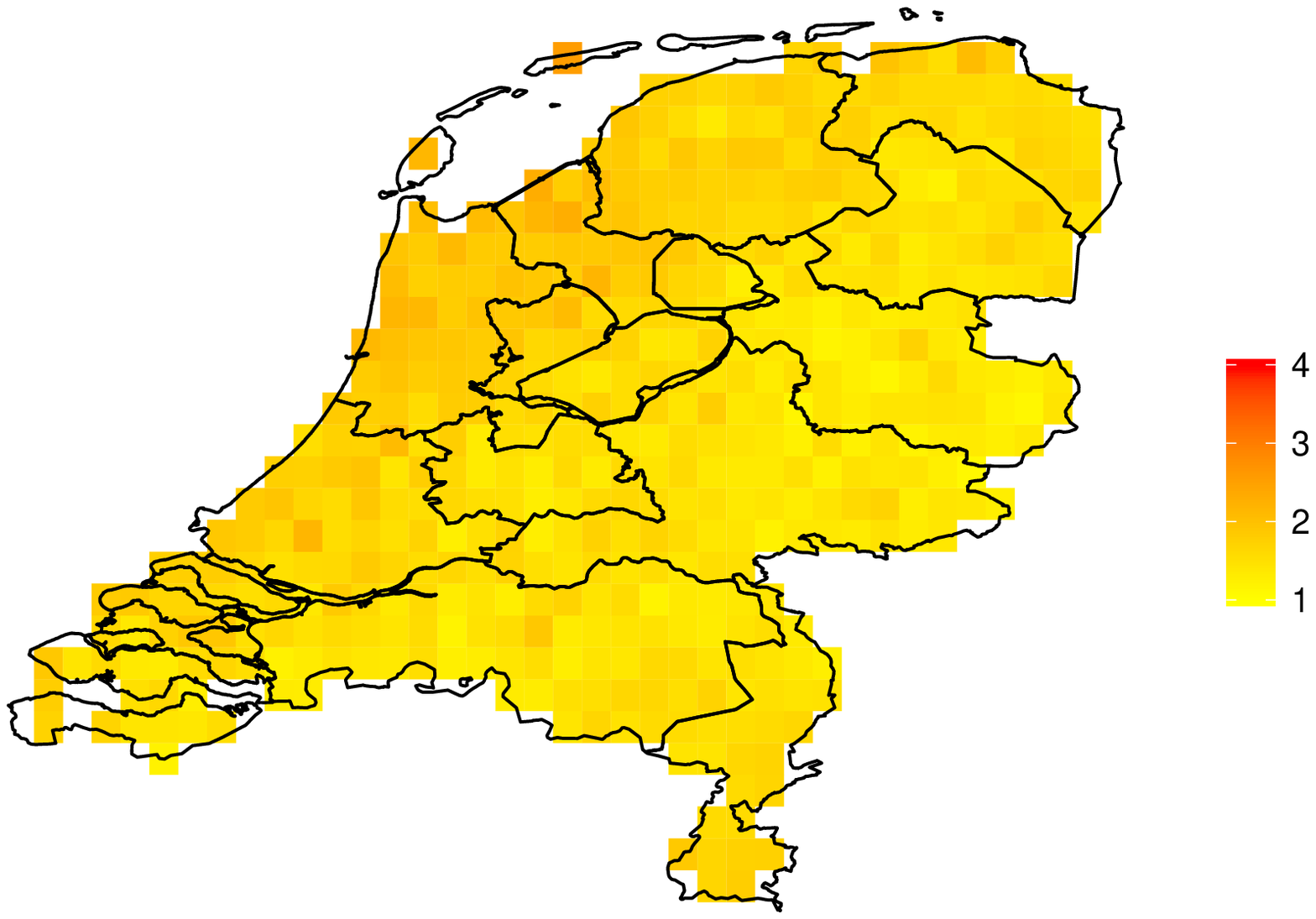}}
\qquad
\subfloat[0.99 quantile, MS]{\includegraphics[width=0.32\linewidth]{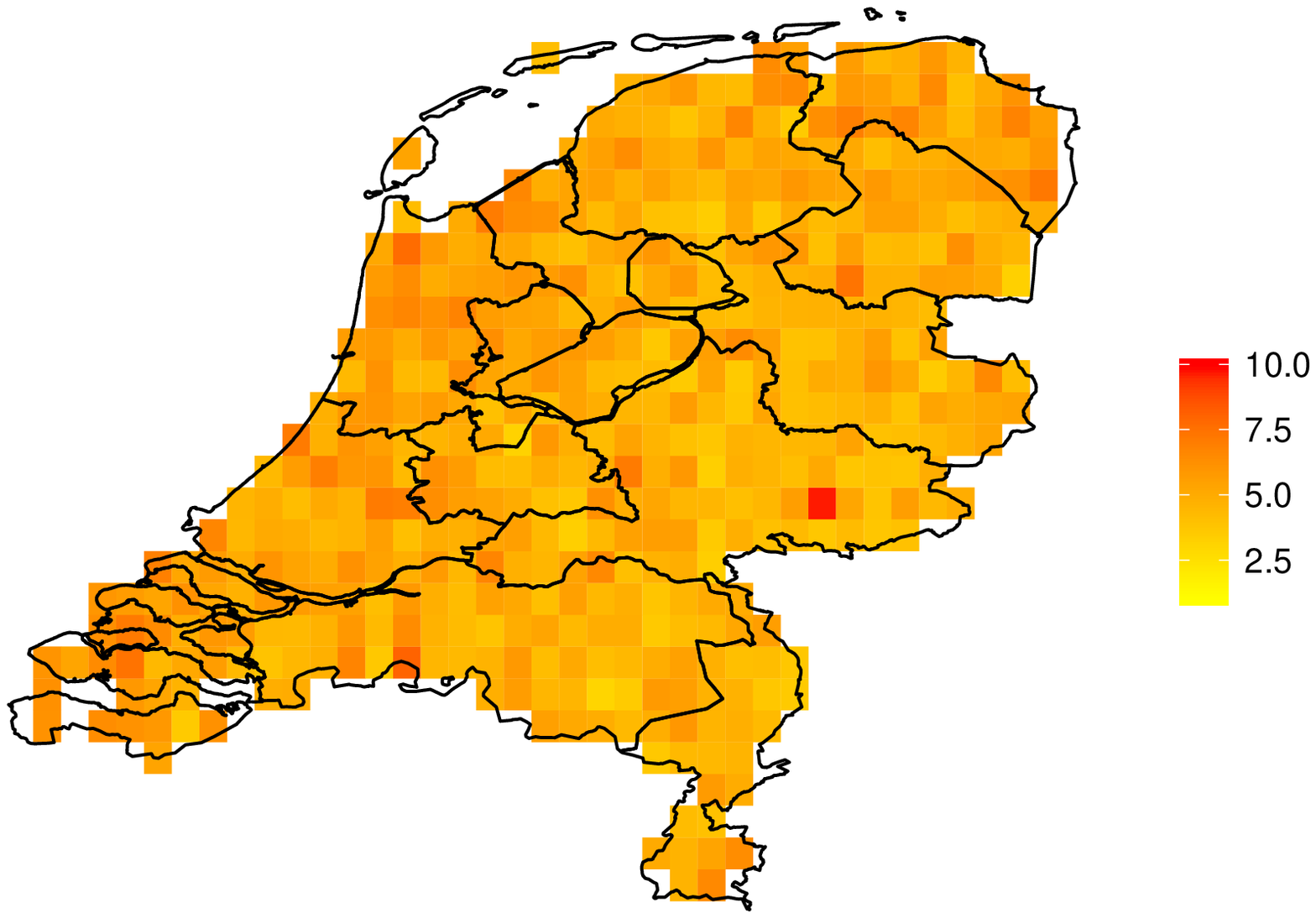}}
\subfloat[0.99 quantile, SB]{\includegraphics[width=0.32\linewidth]{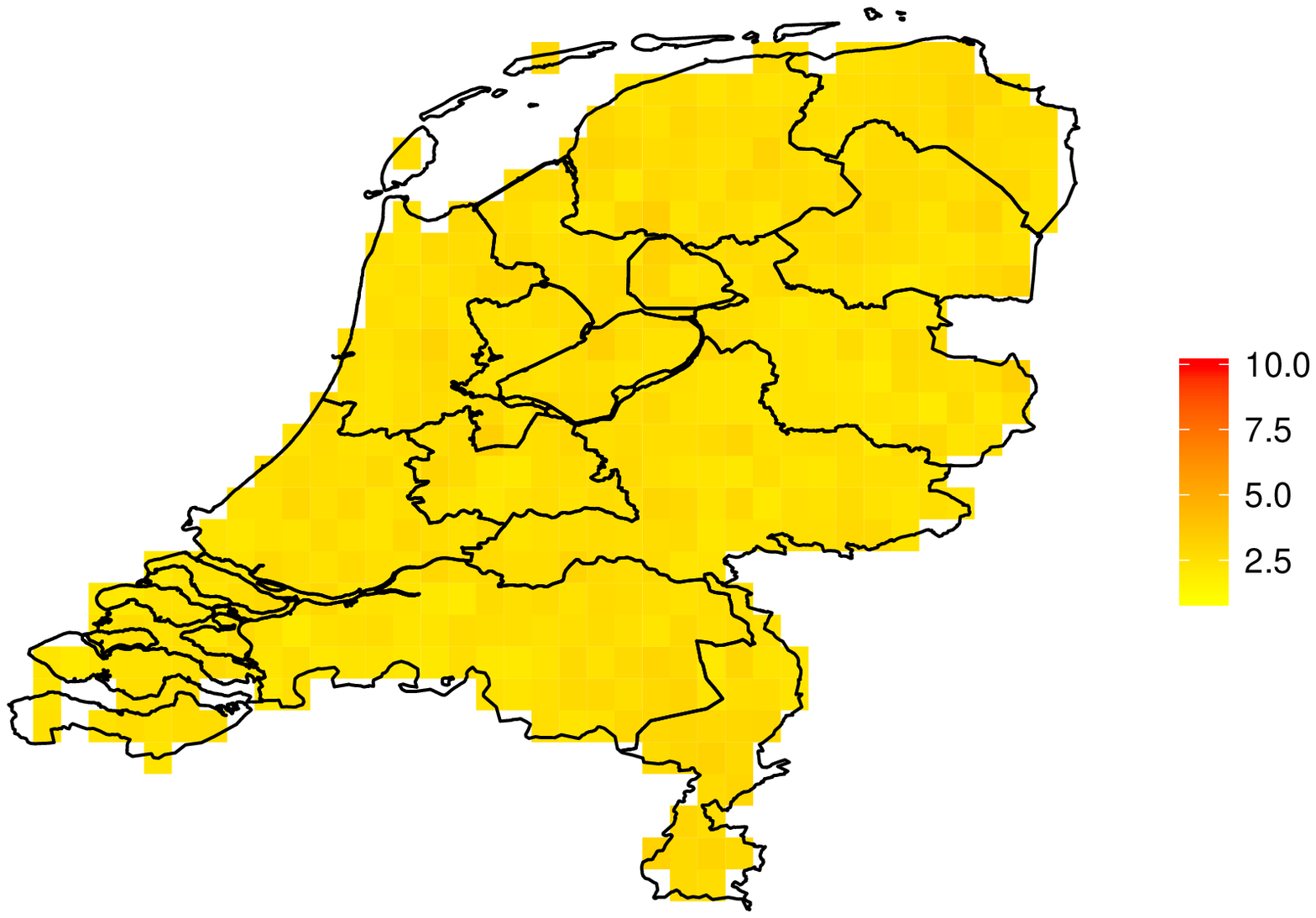}}
\subfloat[0.99 quantile, MM]{\includegraphics[width=0.32\linewidth]{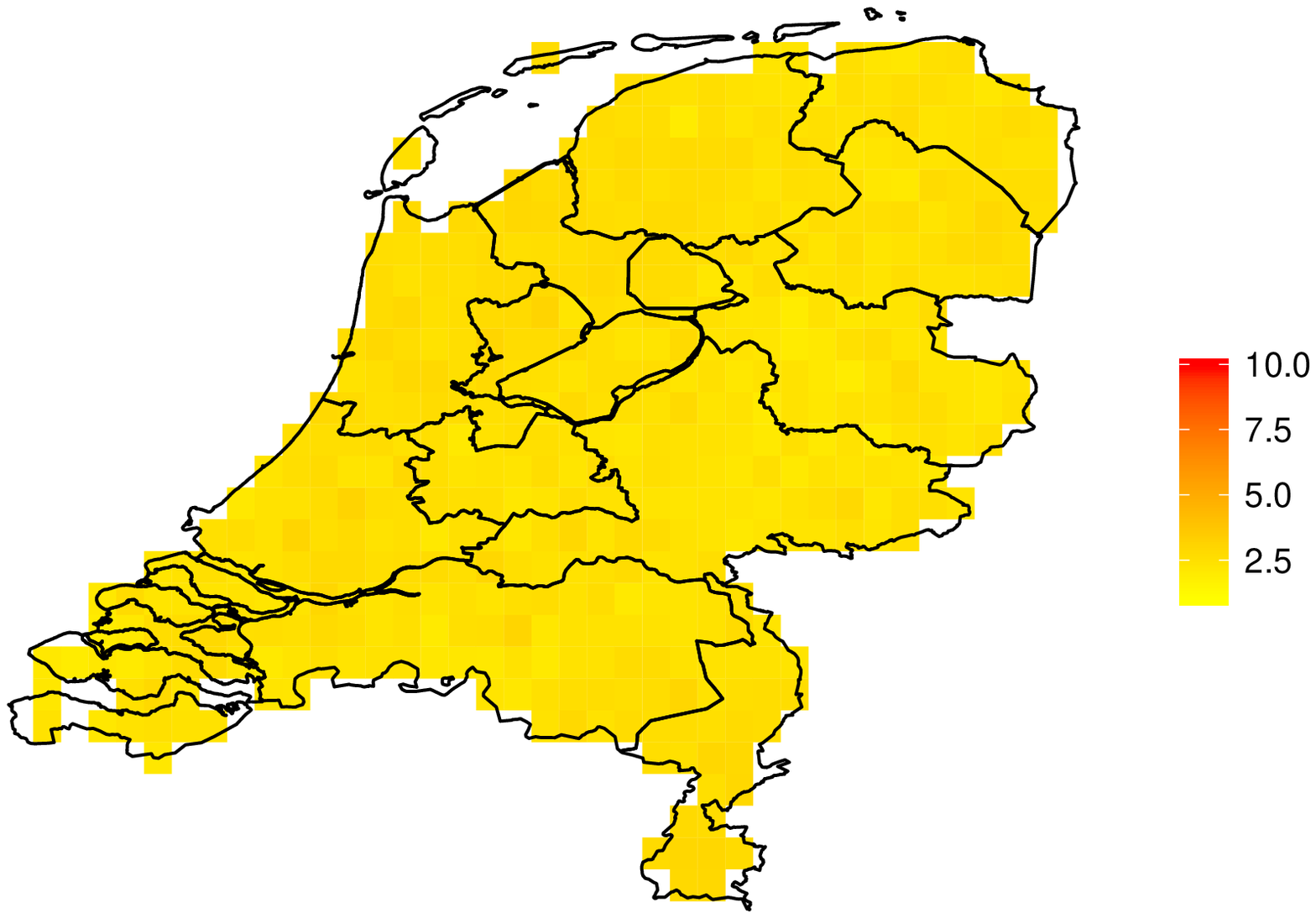}}
\caption{Posterior standard deviation of the 0.50, 0.95 and 0.99 quantiles for the predicted wind speed. All units are m/s.}
	\label{qsd}
\end{figure}

{We follow \cite{reich2012hierarchical} to predict pointwise quantiles at a new location $s^\star$ under model MS, SB and MM. Given historical records of the 30 climate stations, we make predictions for 494 grid cells over the Netherlands. Figures \ref{qmn} and \ref{qsd} plot the posterior mean and standard deviation of various pointwise quantile levels. Overall, the wind speed in the coastal areas and the islands is greater than the inland areas. For the 0.5 quantile level, all three models have similar performances. For higher quantile levels such as 0.95 and 0.99, the MS model tends to make higher predictions than the SB and MM model. This is due to the fact that the MS model has the strongest tail dependence among all these three models. The results of the SB and MM models are close and have smaller pointwise standard deviation than the MS model.}

\section{Discussion}\label{s:discussion}
In this paper, we reviewed the HEVP model and proposed two extensions 
that can span a wider family of spatial processes. 
The SB model generalizes the HEVP model by incorporating a semi-parametric truncated Dirichlet process prior. Unlike the HEVP model that always has tail dependence, the SB model is tail independent. As it is generally hard to identify the tail behavior in practice, we introduce the MM model. As a hybrid of the HEVP model and the SB model, the MM model smoothly bridges the two dependence classes. In the MM model, as we discussed above, $\delta=I\{q\geq \alpha/{(1+\alpha)}\}$ can be treated as an indicator of tail dependence, which shows the potential application on tail dependence test by use of the MM model. The results from the simulation study and data analysis suggest that the MM model generally delivers the more robust performance than the HEVP and SB model. 

The model can be further generalized. For example, in the current SB model, we consider the random effects $\mathbf{A}_t$ on $L$ knots jointly in order to take into account the dependence among these random effects. 
However, nonparametrically estimates this joint distribution requires many replicates of the process. An alternative for case with few replicates is to assume the $A_{lt}$ are independent and identically distributed, and use a Dirichlet process prior for their univariate distribution.
In addition, in the present MM model, we assume a universal tuning parameter $q$ across all quantile levels. In practice, the dependence structures can vary across space. Therefore, we can introduce different $q$ for each partition of the data to allow different contribution of the HEVP and the SB model.

\bibliographystyle{spbasic}      
\bibliography{BSE}   

\section*{Appendix A.1: Computational details}
\label{s:A1}
\subsection*{{A.1.1 SB model}}
We use Metropolis-within-Gibbs MCMC for model fitting and prediction. The details of implementing the HEVP model can be found in \cite{reich2012hierarchical}. In the SB model, the random effect ${\gamma}_{jl}$ follows a positive stable distribution that does not have a closed form. We use the technique in  \cite{stephenson2009high} and incorporate auxiliary variables ${\bm \lambda}_j = (\lambda_{j1},\ldots,\lambda_{jL}) \in [0,1]^L$ such that for each $\lambda_{jl}$,
\[
p(\gamma_{jl},\lambda_{jl}|\alpha) = \frac{\alpha \gamma_{jl}^{-1/(1-\alpha)}}{1-\alpha}c(\lambda_{jl})\exp[-c(\lambda_{jl})\gamma_{jl}^{-\alpha/(1-\alpha)}],
\]
where $c(x)=[\frac{\sin(\alpha\pi x)}{\sin(\pi x)}]^{1/(1-\alpha)}\frac{\sin[(1-\alpha)\pi x]}{\sin(\alpha\pi x)}$. We also introduce the cluster label $g_t \in \{1,\ldots,J\}$ and rewrite the hierarchical representation of the SB model as
\begin{align}
Y_t(s)|\mathbf{A}_{t}, \mathbf{B}_{t}, g_t = j &\stackrel{i.i.d.}{\sim} \mathrm{GEV}\{\mu_{j}^\star(s), \sigma_{j}^\star(s),\xi^\star(s)\},\\
(A_{lt},B_{lt})|g_t=j & \stackrel{i.i.d.}{\sim} p(\gamma_{jl},\lambda_{jl}|\alpha), \nonumber\\
g_t | {\bm \pi} &\stackrel{i.i.d.}{\sim} \mathrm{Cat}({\bm\pi}), \nonumber
\end{align}
{\color{black}where $\mathbf{A}_t=(A_{1t},\ldots,A_{Lt}), \mathbf{B}_t=(B_{1t},\ldots,B_{Lt})$, $\mathrm{Cat}({\bm\pi})$ denotes the categorical distribution with $\mathrm{Prob}(g_t = j) = \pi_j$ for $j=1,\ldots,J$ where ${\bm \pi} = (\pi_1,\ldots,\pi_J)$. 
	More specifically,
	\[\left\{
	\begin{array}{l}
	\mu_{j}^\star(s)=\mu(s)+\frac{\sigma(s)}{\xi(s)}[\theta_{j}(s)^{\xi(s)}-1]\\
	\sigma_{j}^\star(s) = \alpha \sigma(s) \theta_{j}(s)^{\xi(s)}\\
	\xi_{j}^\star(s)= \alpha \xi(s) \\
	\theta_j(s) = [\sum_{l=1}^L\gamma_{jl}\omega_l(s)^{1/\alpha}]^\alpha \\
	\omega_l(s)=\frac{K(s|v_l,\tau)}{\sum_{j=1}^LK(s|v_j,\tau)} \\
	K(s|v_l,\tau)=\frac{1}{2\pi\tau^2}\exp\{-\frac{(s-v_l)^T(s-v_l)}{2\tau^2}\}
	\end{array}
	\right.\]}
Denote $$\bm \Theta=\{\{\mu(s_i)\}_{i=1}^n,\{\sigma(s_i)\}_{i=1}^n,\{\xi(s_i)\}_{i=1}^n,\tau,\alpha, \{{\bm \gamma}_j\}_{j=1}^J,  \{{\bm \lambda}_j\}_{j=1}^J,\{g_t\}_{t=1}^T,{\bm \pi}\}$$ as the set of the model parameters. We follow the update rules in \cite{reich2012hierarchical} to update the GEV parameters $\mu(s_i), \sigma(s_i)$ and $\xi(s_i)$, spatial dependence parameters $\tau$ and $\alpha$. {\color{black}We assume the GEV parameters $\mu(s_i),\log(\sigma(s_i))$ and $\xi(s_i)$ are constant across space and have $N(0,10^2)$, $N(0,1)$ and $N(0,0.25^2)$ priors, respectively. The residual dependence parameters
	have priors $\tau \sim InvGamma(0.1, 0.1)$ and $\alpha \sim Unif(0, 1)$. 
	The auxiliary variables ${\bm \gamma}_j$ and ${\bm \lambda}_j$ are updated using Metropolis sampling. For the $r$th MCMC iteration, we generate a candidate $\gamma_{jl}^{(c)}$ from log-normal distribution $LN(\gamma_{jl}^{(r-1)},s_{\gamma_{jl}}^2)$. Denote $f(\gamma_{jl}^{(c)}|\gamma_{jl}^{(r-1)})$ as the density of the candidate distribution, the acceptance ratio is 
	\begin{align*}
	\left\{ \frac{\prod_{t:g_t=j}\prod_{i=1}^nl[Y_t(s_i)|\theta_j^{(c)},\mathrm{rest}]}{\prod_{t:g_t=j}\prod_{i=1}^nl[Y_t(s_i)|\theta_j^{(r-1)},\mathrm{rest}]}\right\}\left\{\frac{p(\gamma_{jl}^{(c)},\lambda_{jl}|\alpha)}{p(\gamma_{jl}^{(r-1)},\lambda_{jl}|\alpha)}\right\}\left\{\frac{f(\gamma_{jl}^{(c)}|\gamma_{jl}^{(r-1)})}{f(\gamma_{jl}^{(r-1)}|\gamma_{jl}^{(c)})}\right\}.
	\end{align*}
	We use a truncated normal candidate distribution for $\lambda_{jl}\sim TN(\lambda_{jl}^{(r-1)},s_{\lambda_{jl}})$ on $[0,1]$. Denote $f(\lambda_{jl}^{(c)}|\lambda_{jl}^{(r-1)})$ as the density of the candidate function. The acceptance ratio is 
	\begin{align*}
	\left\{\frac{p(\gamma_{jl},\lambda_{jl}^{(c)}|\alpha)}{p(\gamma_{jl},\lambda_{jl}^{(r-1)}|\alpha)}\right\}\left\{\frac{f(\lambda_{jl}^{(c)}|\lambda_{jl}^{(r-1)})}{f(\lambda_{jl}^{(r-1)}|\lambda_{jl}^{(c)})}\right\}
	\end{align*}

}

For $\{\{g_t\}_{t=1}^T,{\bm \pi}\}$, {\color{black}we perform Gibbs sampling from 
	the full conditional posterior distribution}:
\[
Pr(g_t=j|{\rm data,rest}) = \frac{\pi_j\prod_{i=1}^n l[Y_t(s_i)|\mu_j(s_i),\sigma_j(s_i),\xi(s_i),\theta_j(s),\tau, \alpha]}{\sum_{j=1}^J\pi_j\prod_{i=1}^n l[Y_t(s_i)|\mu_j(s_i),\sigma_j(s_i),\xi(s_i),\theta_j(s),\tau, \alpha]}.
\]
Under the stick-breaking representation, $\pi_j=v_j\prod_{i=1}^{j-1}(1-v_i), v_i\stackrel{i.i.d}{\sim} Beta(1,1)$. We update ${\bm \pi}$ by means of updating $(v_1,\ldots,v_J)$, where the posterior distribution of $v_j$ is
\[
v_j | {\rm data,rest} \sim Beta(1+\sum_{t=1}^TI(g_t=j), 1 + \sum_{t=1}^TI(g_t>j)).
\]
\subsection*{{A.1.2 MM model}}
The hierarchical representation of the MM model is
\begin{align}
Y_t(s)|\Tilde{\mathbf{A}}_{t}, \Tilde{\mathbf{B}}_{t}, \hat{\mathbf{A}}_{t}, \hat{\mathbf{B}}_{t}, g_t = j, q &\stackrel{i.i.d.}{\sim} \max\{\mathrm{GEV}\{\Tilde{\mu}_{t}^\star(s), \Tilde{\sigma}_{t}^\star(s),\Tilde{\xi}^\star(s)\},\\
&\,\,\,\,\,\,\,\,\,\,\,\,\,\,\,\,\,\,\mathrm{GEV}\{\hat{\mu}_{j}^\star(s), \hat{\sigma}_{j}^\star(s),\hat{\xi}^\star(s)\}\},\nonumber\\
(\Tilde{A}_{lt},\Tilde{B}_{lt}) & \stackrel{i.i.d.}{\sim} p(A,B|\alpha),\nonumber\\
(\hat{A}_{lt},\hat{B}_{lt})|g_t=j & \stackrel{i.i.d.}{\sim} p(\gamma_{jl},\lambda_{jl}|\alpha), \nonumber\\
g_t | {\bm \pi} &\stackrel{i.i.d.}{\sim} \mathrm{Cat}({\bm\pi}), \nonumber
\end{align}
where
\[\left\{
\begin{array}{l}
\Tilde{\mu}_{t}^\star(s)=\mu(s)+\frac{\sigma(s)}{\xi(s)}[\Tilde{\theta}_{t}(s)^{q{\xi}(s)}q^{\xi(s)}-1]\\
\Tilde{\sigma}_t^\star(s)=\alpha q \sigma(s)\Tilde{\theta}_t(s)^{q\xi(s)}q^{\xi(s)}\\
\Tilde{\xi}^\star(s)= \alpha q \xi(s) \\
\Tilde{\theta}_t(s)=\left[\sum_{l=1}^LA_{lt}\omega_l(s)^{1/\alpha}\right]^\alpha
\end{array}
\right.\]
and 
\[\left\{
\begin{array}{l}
\hat{\mu}_{j}^\star(s)=\mu(s)+\frac{\sigma(s)}{\xi(s)}[\hat{\theta}_{j}(s)^{(1-q){\xi}(s)}(1-q)^{\xi(s)}-1] \\
\hat{\sigma}_j^\star(s)=\alpha (1-q) \sigma(s)\hat{\theta}_j(s)^{(1-q)\xi(s)}(1-q)^{\xi(s)}\\
\hat{\xi}^\star(s)= \alpha (1-q) \xi(s) \\
\hat{\theta}_j(s)=\left[\sum_{l=1}^L\gamma_{jl}\omega_l(s)^{1/\alpha}\right]^\alpha
\end{array}
\right.\]
and  $\omega_l(s)=\frac{K(s|v_l,\tau)}{\sum_{j=1}^LK(s|v_j,\tau)}$,  $K(s|v_l,\tau)=\frac{1}{2\pi\tau^2}\exp\{-\frac{(s-v_l)^T(s-v_l)}{2\tau^2}\}$.

{\color{black}Denote 
	$$\bm \Delta=\{\bm \Theta, \{\Tilde{{\bm A}}_t\}_{t=1}^T, \{\Tilde{{\bm B}}_t\}_{t=1}^T, q\}$$ as the set of the model parameters.
	We follow A1.1 and \cite{reich2012hierarchical} to choose the priors and updating rules for the parameters in $\bm \Theta$ and $\{\{\Tilde{{\bm A}}_t\}_{t=1}^T, \{\Tilde{{\bm B}}_t\}_{t=1}^T\}$, respectively.}
The tuning paramter $q$ is updated using Metropolis sampling. For the $r$th MCMC iteration, we generate a candidate $q^{(c)}$ from truncated normal distribution $TN(q^{(r-1)}, s_q^2)$ on $[0,1]$, where $q^{(r-1)}$ is the value at the $(r-1)$th iteration and $s_q$ is the tuning paramter. We choose $U[0,1]$ prior for $q$, the acceptance ratio is 
\[
\left\{\frac{\prod_{t=1}^T\prod_{i=1}^n l[Y_t(s_i)|q^{(c)},\Tilde{\theta}^{(c)}_t(s),\hat{\theta}^{(c)}_j(s),{\rm rest}]}{\prod_{t=1}^T\prod_{i=1}^n l[Y_t(s_i)|q^{(r-1)},\Tilde{\theta}^{(r-1)}_t(s),\hat{\theta}^{(r-1)}_j(s),{\rm rest}]}\right\}\left\{\frac{f(q^{(c)}|q^{(r-1)})}{f(q^{(r-1)}|q^{(c)})}\right\}.
\]

\section*{Appendix A.2: Derivations}
\subsection*{{ A.2.1 Derivation of tail behavior for the stick-breaking prior model }}\label{s:A2}
Consider the SB model, the marginal cumulative distribution function at site $s$ is
$$F_{SB}(x;s|\bm \gamma) =  \sum_{j=1}^{J}\pi_j\exp\left\{-\left[x^{-1/\alpha}\sum_{l=1}^{L}
\omega_l(s)^{1/\alpha} \gamma_{lj}\right]\right\}.$$
\textit{Tail index}:\\
Using the Taylor series expansion of the exponential function we have
$1-F_{SB}(x;s|\bm \gamma)\sim \sum_{j=1}^{J}\pi_j\left[x^{-1/\alpha}\sum_{l=1}^{L}
\omega_l(s)^{1/\alpha} \gamma_{lj}\right] (x\to \infty)$.
Therefore the tail index is 
\begin{align*}
a_{SB} &= \liminf_{x\to\infty}\frac{-\log\{1-F_{SB}(x;s|\bm \gamma)\}}{\log(x)} \\
&= \lim_{x\to\infty}\frac{-\log\left[\sum_{j=1}^{J}\pi_j\left\{x^{-1/\alpha}\sum_{l=1}^{L} \omega_l(s)^{1/\alpha} \gamma_{lj}\right\}\right]}{\log(x)}\nonumber \allowdisplaybreaks\\
&=\liminf_{x\to\infty}\frac{-\log\left(\left[\sum_{j=1}^{J}\pi_j\left\{\sum_{l=1}^{L} \omega_l(s)^{1/\alpha} \gamma_{lj}\right\}\right]x^{-1/\alpha}\right)}{\log(x)}\nonumber\\
&= \liminf_{x\to\infty}\frac{-\log(x^{-1/\alpha})-\log\left(\left[\sum_{j=1}^{J}\pi_j\left\{\sum_{l=1}^{L} \omega_l(s)^{1/\alpha} \gamma_{lj}\right\}\right]\right)}{\log(x)} \nonumber\\
&= 1/\alpha.
\end{align*}
\noindent\textit{Tail dependence}:\\
According to Jensen's inequality, we have
$$F_{SB}(x;s|\bm \gamma) \geq \exp\left\{-\sum_{j=1}^J\pi_j\left[\sum_{l=1}^{L}
\omega_l(s)^{1/\alpha} \gamma_{lj}\right]x^{-1/\alpha}\right\}.$$ Take inverse on the right side, we have 
\begin{align*}
F_{SB}^{-1}(u;s|\bm \gamma) \leq \left(-\frac{\sum_{j=1}^J\pi_j\left[\sum_{l=1}^{L}
	\omega_l(s)^{1/\alpha} \gamma_{lj}\right]}{\log u} \right)^{\alpha}.
\end{align*}
The pairwise tail dependence for site $s_1$ and $s_2$ is defined as $$\chi_{SB}(s_1,s_2)=\lim\limits_{u\to 1}P\{X_t(s_1)>F^{-1}_{SB}(u;s_1|\bm \gamma)|X_t(s_2)>F^{-1}_{SB}(u;s_2|\bm \gamma),\bm \gamma\}.$$ 
It is straightforward to show that
\begin{equation}\label{chi}
\chi_{SB}(s_1,s_2) = \lim\limits_{u\to 1}\frac{1-2u+P\{X_t(s_1)\leq F^{-1}_{SB}(u;s_1|\bm \gamma), X_t(s_2)\leq F^{-1}_{SB}(u;s_2|\bm\gamma)|\bm \gamma\}}{1-u}.
\end{equation}
According to \eqref{eq8}, we have
\begin{align*}
&P\{X_t(s_1)\leq F^{-1}_{SB}(u;s_1|\bm\gamma), X_t(s_2)\leq F^{-1}_{SB}(u;s_2|\bm\gamma)|\bm \gamma\} \\
&= \sum_{j=1}^{J}\pi_j\exp\left\{-\left[\sum_{l=1}^L \left(
\left\{\frac{\omega_l(s_1)}{F^{-1}_{SB}(u;s_1|\bm\gamma)}\right\}^{1/\alpha}+\left\{\frac{\omega_l(s_2)}{F^{-1}_{SB}(u;s_2|\bm\gamma)}\right\}^{1/\alpha} \right)\gamma_{lj} \right] \right\}\allowdisplaybreaks\\
&\leq \sum_{j=1}^{J}\pi_j\exp\left\{\left[ \frac{\sum_{l=1}^L\omega_l(s_1)^{1/\alpha}\gamma_{lj}}{\sum_{j=1}^J\pi_j\sum_{l=1}^L\omega_l(s_1)^{1/\alpha}\gamma_{lj}}+\frac{\sum_{l=1}^L\omega_l(s_2)^{1/\alpha}\gamma_{lj}}{\sum_{j=1}^J\pi_j\sum_{l=1}^L\omega_l(s_2)^{1/\alpha}\gamma_{lj}}\right]\left(\log u\right)\right\} \\ 
&=\sum_{j=1}^J \pi_j u^{\left[ \frac{\sum_{l=1}^L\omega_l(s_1)^{1/\alpha}\gamma_{lj}}{\sum_{j=1}^J\pi_j\sum_{l=1}^L\omega_l(s_1)^{1/\alpha}\gamma_{lj}}+\frac{\sum_{l=1}^L\omega_l(s_2)^{1/\alpha}\gamma_{lj}}{\sum_{j=1}^J\pi_j\sum_{l=1}^L\omega_l(s_2)^{1/\alpha}\gamma_{lj}}\right]}.
\end{align*}
Hence 
\begin{align*}
\chi_{SB}(s_1,s_2) &\leq \lim\limits_{u\to 1}\frac{1-2u+\sum_{j=1}^J \pi_j u^{\left[ \frac{\sum_{l=1}^L\omega_l(s_1)^{1/\alpha}\gamma_{lj}}{\sum_{j=1}^J\pi_j\sum_{l=1}^L\omega_l(s_1)^{1/\alpha}\gamma_{lj}}+\frac{\sum_{l=1}^L\omega_l(s_2)^{1/\alpha}\gamma_{lj}}{\sum_{j=1}^J\pi_j\sum_{l=1}^L\omega_l(s_2)^{1/\alpha}\gamma_{lj}}\right]}}{1-u}\\
&= \sum_{j=1}^J\pi_j \left[ \frac{\sum_{l=1}^L\omega_l(s_1)^{1/\alpha}\gamma_{lj}}{\sum_{j=1}^J\pi_j\sum_{l=1}^L\omega_l(s_1)^{1/\alpha}\gamma_{lj}}+\frac{\sum_{l=1}^L\omega_l(s_2)^{1/\alpha}\gamma_{lj}}{\sum_{j=1}^J\pi_j\sum_{l=1}^L\omega_l(s_2)^{1/\alpha}\gamma_{lj}}\right] - 2\\
&= 2-2 = 0.
\end{align*}
As $\chi_{SB}(s_1,s_2)\geq 0$ by definition, therefore $\chi_{SB}(s_1,s_2)=0$.

\subsection*{A.2.2 Derivation of tail dependence for max-mixture hybrid model}\label{s:A22}
Denote $F_{MM}(x;s|\bm\gamma)=G_1(x)G_2(x;s|\bm\gamma)$, where
\begin{align}
\nonumber
G_1(x)&=\exp(-q^{\frac{1}{q}}x^{-\frac{1}{q}}) \nonumber \\
G_2(x;s|\bm\gamma)&=\sum_{j=1}^{J}\pi_j\exp\left\{-\left[\sum_{l=1}^{L}
{\omega_l(s)}^{\frac{1}{\alpha}} \gamma_{lj}\right](1-q)^{\frac{1}{(1-q)\alpha}}x^{-\frac{1}{(1-q)\alpha}}\right\} \nonumber,
\end{align}
we have $1-G_1(x) \sim q^{\frac{1}{q}}x^{-\frac{1}{q}} (x\to \infty)$ and $1-G_2(x;s|\bm\gamma) \sim \sum_{j=1}^J\pi_j\left[\sum_{l=1}^{L}{\omega_l(s)}^{\frac{1}{\alpha}} \gamma_{lj}\right](1-q)^{\frac{1}{(1-q)\alpha}}x^{-\frac{1}{(1-q)\alpha}} (x\to \infty)$. 

Similarly we can show that
\begin{equation}
\chi_{MM}(s_1,s_2) = \lim\limits_{u\to 1}\frac{1-2u+P\{X_t(s_1)\leq F^{-1}_{MM}(u;s_1|\bm\gamma), X_t(s_2)\leq F^{-1}_{MM}(u;s_2|\bm\gamma)|\bm\gamma\}}{1-u}. \nonumber
\end{equation}
\textit{Case 1}: $q>\frac{\alpha}{1+\alpha}$: 
\begin{itemize}
	\item[(i)] By definition, $1-F_{MM}(x;s|\bm\gamma)\geq 1-G_1(x)$, therefore $F_{MM}^{-1}(u;s|\bm\gamma)\geq q[-\log(u)]^{-q}$. Derive the third term in the numerator of \eqref{chi} by \eqref{eq666}, we have
	\begin{align}\label{eq54}
	&P\{X_t(s_1)\leq F^{-1}_{MM}(u;s_1|\bm\gamma), X_t(s_2)\leq F^{-1}_{MM}(u;s|\bm\gamma)|\bm\gamma\} \\
	&\geq \exp\left\{-\sum_{l=1}^{L}\left(
	\omega_l(s_1)^{1/\alpha} + \omega_l(s_2)^{1/\alpha}\right)^\alpha\left({-\log u}\right)\right\} \nonumber \\
	&\times \sum_{j=1}^{J}\pi_j \exp\left\{-\left[\sum_{l=1}^{L}\left(
	\omega_l(s_1)^{1/\alpha} + \omega_l(s_2)^{1/\alpha}\right)\gamma_{jl} \right]
	\left({-\log u}\right)^{\frac{q}{(1-q)\alpha}} \right\}\nonumber \allowdisplaybreaks\\
	&=u^{\sum_{l=1}^{L}\left(\omega_l(s_1)^{1/\alpha} + \omega_l(s_2)^{1/\alpha}\right)^\alpha} \nonumber\\
	&\times \sum_{j=1}^{J}\pi_j \exp\left\{-\left[\sum_{l=1}^{L}\left(
	\omega_l(s_1)^{1/\alpha} + \omega_l(s_2)^{1/\alpha}\right)\gamma_{jl} \right]\left(\frac{1-q}{q}\right)^{\frac{1}{(1-q)\alpha}}
	\left({-\log u}\right)^{\frac{q}{(1-q)\alpha}} \right\}\nonumber\\
	&=T_1\times T_2,\nonumber
	\end{align}
	where 
	\begin{align}
	T_1 &= u^{\sum_{l=1}^{L}\left(\omega_l(s_1)^{1/\alpha} + \omega_l(s_2)^{1/\alpha}\right)^\alpha} \nonumber\\
	T_2 &= \sum_{j=1}^{J}\pi_j \exp\left\{-\left[\sum_{l=1}^{L}\left(
	\omega_l(s_1)^{1/\alpha} + \omega_l(s_2)^{1/\alpha}\right)\gamma_{jl} \right]\left(\frac{1-q}{q}\right)^{\frac{1}{(1-q)\alpha}}
	\left({-\log u}\right)^{\frac{q}{(1-q)\alpha}} \right\}. \nonumber
	\end{align}
	Therefore,
	\begin{equation}
	\chi_{MM}(s_1,s_2) \geq \lim\limits_{u\to 1}\frac{1-2t+T_1}{1-u} + \lim\limits_{u\to 1}\frac{T_1(T_2-1)}{1-u}. \nonumber
	\end{equation}
	The first term on RHS is equal to $2 - \sum_{l=1}^{L}(\omega_l(s_1)^{1/\alpha}+\omega_l(s_2)^{1/\alpha})^{\alpha}$. Use L'Hospital for the second term on RHS, we have
	\begin{align} \label{eq55}
	\lim\limits_{u\to 1}\frac{T_1(T_2-1)}{1-u} &= \lim\limits_{u\to 1}\left\{\sum_{j=1}^{J}\pi_j\left[\sum_{l=1}^{L}\left(
	\omega_l(s_1)^{1/\alpha} + \omega_l(s_2)^{1/\alpha}\right)\right]\gamma_{jl}\left(\frac{1-q}{q}\right)^{\frac{1}{(1-q)\alpha}}\right\}\\
	&\times
	\exp\left\{-\left[\sum_{l=1}^{L}\left(
	\omega_l(s_1)^{1/\alpha} + \omega_l(s_2)^{1/\alpha}\right)\left({-\log u} \right)^{\frac{q}{(1-q)\alpha}} \left(\frac{1-q}{q}\right)^{\frac{1}{(1-q)\alpha}}\gamma_{jl} \right] \right\} \nonumber\\
	&\times \frac{q}{(1-q)\alpha}\left(-\log u\right)^{\frac{q}{(1-q)\alpha}-1}\frac{1}{u} \nonumber\\
	&=0.\nonumber
	\end{align}
	Therefore, $\chi_{MM}(s_1,s_2)\geq 2 - \sum_{l=1}^{L}(\omega_l(s_1)^{1/\alpha}+\omega_l(s_2)^{1/\alpha})^{\alpha}$.
	
	\item[(ii)] Notice that for any positive $\epsilon$ there exists $N$ such that $1-G_2(x;s|\bm\gamma)\leq\epsilon(1-G_1(x))$ for all $x>N$. Therefore
	\[
	1-F_{MM}(x;s|\bm\gamma) = 1-G_2(x;s|\bm\gamma) + G_2(x;s|\bm\gamma)[1-G_1(x)] \leq (1+\epsilon)[1-G_1(x)].
	\]
	It's straightforward to show that $F_{MM}^{-1}(u;s|\bm\gamma)\leq q\left[-\log\left(\frac{u+\epsilon}{1+\epsilon}\right)\right]^{-q}$.
	Consequently,
	\begin{align*}
	&P\{X_t(s_1)\leq F^{-1}_{MM}(u;s_1|\bm\gamma), X_t(s_2)\leq F^{-1}_{MM}(u;s_2|\bm\gamma)\} \\
	&\leq \exp\left\{-\sum_{l=1}^{L}\left(
	\omega_l(s_1)^{1/\alpha} + \omega_l(s_2)^{1/\alpha}\right)^\alpha\left[{-\log \left(\frac{u+\epsilon}{1+\epsilon}\right)}\right]\right\}\\
	&=\left(\frac{u+\epsilon}{1+\epsilon}\right)^{\sum_{l=1}^{L}\left(\omega_l(s_1)^{1/\alpha} + \omega_l(s_2)^{1/\alpha}\right)^\alpha}.
	\end{align*}
	By L'Hospital we have
	\(
	\chi_{MM}(s_1,s_2) \leq 2 - \frac{1}{1+\epsilon} \sum_{l=1}^{L}(\omega_l(s_1)^{1/\alpha}+\omega_l(s_2)^{1/\alpha})^{\alpha}.
	\)
	Since $\epsilon$ can be chosen arbitrarily small, we've shown $\chi_{MM}(s_1,s_2) \leq 2 - \sum_{l=1}^{L}(\omega_l(s_1)^{1/\alpha}+\omega_l(s_2)^{1/\alpha})^{\alpha}$.
\end{itemize}
(i) and (ii) together yields $\chi_{MM}(s_1,s_2) = 2 - \sum_{l=1}^{L}(\omega_l(s_1)^{1/\alpha}+\omega_l(s_2)^{1/\alpha})^{\alpha}$ under \textit{Case 1}.

\noindent\textit{Case 2}: $q<\frac{\alpha}{1+\alpha}$: 

Notice that for any positive $\epsilon$ there exists $N$ such that $1-G_1(x)\leq\epsilon(1-G_2(x))$ for all $x>N$. Therefore
\[
1-F_{MM}(x;s|\bm\gamma) = 1-G_1(x) + G_1(x)[1-G_2(x;s|\bm\gamma)] \leq (1+\epsilon)[1-G_2(x;s|\bm\gamma)].
\]
It's straightforward to show that $$F_{MM}^{-1}(u;s|\bm\gamma) \leq (1-q) \left(-\frac{\sum_{j=1}^J\pi_j\left[\sum_{l=1}^{L}
	\omega_l(s)^{1/\alpha} \gamma_{lj}\right]}{\log (\frac{u+\epsilon}{1+\epsilon})} \right)^{(1-q)\alpha}.$$
Thus
\begin{align*}
&P\{X_t(s_1)\leq F^{-1}_{MM}(u;s_1|\bm\gamma), X_t(s_2)\leq F^{-1}_{MM}(u;s_2|\bm\gamma)|\bm\gamma\} \\
&=\sum_{j=1}^J \pi_j \left(\frac{u+\epsilon}{1+\epsilon}\right)^{\left[ \frac{\sum_{l=1}^L\omega_l(s_1)^{1/\alpha}\gamma_{lj}}{\sum_{j=1}^J\pi_j\sum_{l=1}^L\omega_l(s_1)^{1/\alpha}\gamma_{lj}}+\frac{\sum_{l=1}^L\omega_l(s_2)^{1/\alpha}\gamma_{lj}}{\sum_{j=1}^J\pi_j\sum_{l=1}^L\omega_l(s_2)^{1/\alpha}\gamma_{lj}}\right]}
\nonumber
\end{align*}
By L'Hospital we have
$\chi_{MM}(s_1,s_2) \leq \frac{2\epsilon}{1+\epsilon}$. Since $\epsilon$ can be chosen arbitrarily small, we've shown $\chi_{MM}(s_1,s_2)=0$ under \textit{Case 2}.

\noindent \textit{Case 3}: $q=\frac{\alpha}{1+\alpha}$:

Similarly to the derivation in \textit{Case 1} (i), we have $\chi_{MM}(s_1,s_2)\geq 2 - \sum_{l=1}^{L}(\omega_l(s_1)^{1/\alpha}+\omega_l(s_2)^{1/\alpha})^{\alpha}.$ On the other hand, $\chi_{MM}(s_1,s_2)\leq1$ by definition. Therefore, we have $$2 - \sum_{l=1}^{L}(\omega_l(s_1)^{1/\alpha}+\omega_l(s_2)^{1/\alpha})^{\alpha}\leq \chi_{MM}(s_1,s_2)\leq 1$$ under \textit{Case 3}.

\subsection*{A.3 Additional results}\label{s:A3}
The values defined in \eqref{value} are presented in the following tables and associated standard errors are given in the parentheses.
\begin{table}[ht]
	\centering
	\caption{$\log\left[\mathrm{MMSE}(\hat{Q}_{\kappa})\right]$ for quantile levels $\tau=0.1,0.3,\ldots,0.9,0.95$, $0.99,0.995$. Standard errors are given in parentheses. ``HEVP" stands for the hierachical extreme value process model, ``SB" stands for the extended stick-breaking prior model, ``MM" stands for the max-mixture hybrid model.}
	\setlength{\tabcolsep}{2pt} 
	\begin{tabular}{c|c|cccccccc}
		\toprule
		Setting& Model & 0.1 & 0.3 & 0.5 & 0.7 & 0.9 & 0.95 & 0.99 & 0.995 \\ 
		\midrule
		\multirow{6}{*}{MS}& HEVP & -6.42 & -6.08 & -5.72 & -5.25 & -4.10 & -3.31 & -1.65 & -1.02 \\ 
		& & (0.17) & (0.18) & (0.17) & (0.17) & (0.15) & (0.18) & (0.24) & (0.26) \\ 
		&  SB & -3.15 & -3.19 & -2.78 & -2.14 & -0.90 & 0.13 & 1.88 & 2.42 \\
		& & (0.06) & (0.07) & (0.07) & (0.08) & (0.07) & (0.06) & (0.03) & (0.03) \\
		&  MM & -6.25 & -5.04 & -4.24 & -3.52 & -2.54 & -2.08 & -1.15 & -0.78 \\ 
		&& (0.17) & (0.19) & (0.17) & (0.16) & (0.16) & (0.15) & (0.15) & (0.14) \\
		\hline
		\multirow{6}{*}{SB}&HEVP & -3.13 & -2.41 & -1.94 & -1.26 & -0.62 & -0.08 & 1.16 & 1.61 \\
		&& (0.01) & ($<$0.01) & (0.01) & ($<$0.01) & (0.02) & (0.03) & (0.04) & (0.04) \\
		&  SB & -3.99 & -3.46 & -2.81 & -2.54 & -1.99 & -1.50 & -1.37 & -1.35 \\ 
		&& (0.16) & (0.17) & (0.14) & (0.13) & (0.10) & (0.08) & (0.07) & (0.07) \\
		&  MM & -4.09 & -3.25 & -2.72 & -2.48 & -1.87 & -1.44 & -1.32 & -1.29 \\
		&& (0.17) & (0.24) & (0.20) & (0.11) & (0.12) & (0.08) & (0.08) & (0.08) \\ 
		\hline
		\multirow{6}{*}{GP}& HEVP & -5.79 & -5.89 & -5.13 & -3.90 & -2.52 & -2.08 & -1.62 & -1.57 \\ 
		&&(0.19) & (0.19) & (0.15) & (0.10) & (0.06) & (0.06) & (0.07) & (0.07) \\
		&  SB & -3.71 & -3.87 & -3.87 & -3.88 & -4.01 & -4.11 & -4.37 & -4.45 \\ 
		&& (0.12) & (0.13) & (0.13) & (0.12) & (0.11) & (0.11) & (0.10) & (0.11) \\
		&  MM & -4.57 & -4.69 & -4.69 & -4.61 & -4.43 & -4.35 & -4.24 & -4.10 \\ 
		&& (0.07) & (0.06) & (0.06) & (0.08) & (0.13) & (0.19) & (0.36) & (0.39) \\ 
		\hline
		\multirow{6}{*}{ST}& HEVP & -3.79 & -3.13 & -2.71 & -2.39 & -1.83 & -1.46 & -0.58 & -0.18 \\
		&& (0.16) & (0.16) & (0.16) & (0.17) & (0.19) & (0.21) & (0.22) & (0.22) \\
		&  SB & -4.34 & -3.99 & -3.58 & -2.75 & -1.24 & -0.42 & 1.03 & 1.50 \\ 
		&& (0.15) & (0.21) & (0.16) & (0.13) & (0.13) & (0.10) & (0.07) & (0.06) \\
		&  MM & -4.22 & -3.71 & -3.09 & -2.32 & -1.19 & -0.67 & 0.38 & 0.83 \\
		&& (0.19) & (0.16) & (0.17) & (0.17) & (0.15) & (0.15) & (0.13) & (0.12) \\
		\hline
		\multirow{6}{*}{InvMS}&HEVP & -3.68 & -5.46 & -4.51 & -2.55 & -0.71 & -0.11 & 0.69 & 0.90 \\ 
		& & (0.16) & (0.19) & (0.14) & (0.06) & (0.04) & (0.04) & (0.04) & (0.04) \\ 
		&  SB & -0.91 & -3.03 & -3.35 & -3.48 & -3.23 & -3.13 & -3.38 & -3.59 \\
		& & (0.03) & (0.08) & (0.09) & (0.09) & (0.08) & (0.08) & (0.08) & (0.09) \\
		&  MM & -3.13 & -3.37 & -3.61 & -3.94 & -4.22 & -4.09 & -3.25 & -2.90 \\
		&& (0.11) & (0.08) & (0.06) & (0.06) & (0.06) & (0.14) & (0.33) & (0.35) \\
		\hline
		\multirow{6}{*}{MAX}&HEVP & -0.10 & 0.15 & 0.52 & 1.03 & 1.91 & 2.38 & 3.36 & 3.74 \\
		&& (0.04) & (0.05) & (0.12) & (0.16) & (0.19) & (0.20) & (0.22) & (0.24) \\ 
		&  SB & -4.86 & -5.29 & -4.86 & -4.09 & -3.12 & -2.66 & -1.42 & -0.77 \\
		&& (0.06) & (0.08) & (0.09) & (0.09) & (0.07) & (0.10) & (0.09) & (0.05) \\ 
		&  MM & -5.07 & -6.29 & -6.38 & -5.79 & -3.76 & -2.93 & -1.66 & -1.21 \\
		&& (0.06) & (0.06) & (0.08) & (0.11) & (0.08) & (0.08) & (0.06) & (0.07) \\
		\bottomrule
	\end{tabular}
\end{table}

\begin{table}[ht]
	\centering
	\caption{$\log\left[\mathrm{MMSE}(\hat{\chi}_u)\times1000\right]$ for quantile levels $\tau=0.1,0.2,\ldots,0.9, 0.95$, $0.99,0.995$. Standard errors are given in parentheses. ``HEVP" stands for the hierachical extreme value process model, ``SB" stands for the extended stick-breaking prior model, ``MM" stands for the max-mixture hybrid model.}
	\setlength{\tabcolsep}{2pt} 
	\begin{tabular}{c|c|cccccccc}
		\toprule
		Setting& Model & 0.1 & 0.3 & 0.5 & 0.7 & 0.9 & 0.95 & 0.99 & 0.995 \\ 
		\midrule
		\multirow{6}{*}{MS}&HEVP & -8.75 & -6.03 & -4.78 & -3.96 & -3.35 & -3.22 & -3.12 & -3.11 \\ 
		& & (0.19) & (0.19) & (0.20) & (0.20) & (0.20) & (0.20) & (0.20) & (0.20) \\
		&  SB & -3.11 & -0.22 & 1.09 & 1.89 & 3.13 & 3.61 & 3.95 & 3.99 \\ 
		&&(0.02) & (0.04) & (0.04) & (0.03) & (0.01) & (0.01) & ($<$0.01) & ($<$0.01) \\
		&  MM & -6.66 & -4.01 & -2.79 & -2.00 & -1.41 & -1.28 & -1.19 & -1.18 \\
		&&(0.16) & (0.16) & (0.16) & (0.16) & (0.16) & (0.16) & (0.16) & (0.16) \\ 
		\hline
		\multirow{6}{*}{SB}& HEVP & -2.22 & 1.27 & 2.90 & 3.72 & 2.86 & 2.92 & 3.19 & 3.23 \\
		&& ($<$0.01) & ($<$0.01) & ($<$0.01) & ($<$0.01) & (0.01) & (0.02) & (0.02) & (0.01) \\
		&  SB & -3.22 & 0.31 & 1.81 & 2.69 & 2.58 & 1.51 & -1.68 & -3.09 \\ 
		&& (0.10) & (0.11) & (0.11) & (0.08) & (0.07) & (0.06) & (0.06) & (0.06) \\
		&  MM & -3.20 & 0.23 & 1.73 & 2.64 & 2.46 & 1.34 & -1.85 & -3.26 \\
		&& (0.17) & (0.12) & (0.12) & (0.09) & (0.06) & (0.06) & (0.06) & (0.06) \\
		\hline
		\multirow{6}{*}{GP}&HEVP & -5.87 & -4.03 & -3.17 & -1.19 & 0.90 & 1.49 & 2.13 & 2.26 \\
		&& (0.07) & (0.10) & (0.12) & (0.11) & (0.06) & (0.05) & (0.03) & (0.03) \\
		&  SB & -4.14 & -1.40 & -0.31 & 0.22 & 0.22 & -0.05 & -1.14 & -1.73 \\
		&& (0.13) & (0.09) & (0.08) & (0.07) & (0.04) & (0.02) & (0.01) & (0.01) \\ 
		&  MM & -4.79 & -2.04 & -0.90 & -0.29 & -0.11 & -0.31 & -0.86 & -0.96 \\ 
		&& (0.03) & (0.03) & (0.03) & (0.03) & (0.04) & (0.04) & (0.27) & (0.41) \\ 
		\hline
		\multirow{6}{*}{ST}&HEVP & -2.42 & 1.19 & 3.06 & 3.95 & 4.39 & 4.40 & 4.28 & 4.22 \\
		&& (0.10) & (0.07) & (0.05) & (0.05) & (0.06) & (0.06) & (0.07) & (0.07) \\ 
		& SB & -2.09 & -0.12 & 2.26 & 3.43 & 4.71 & 5.35 & 5.77 & 5.79 \\
		& & (0.07) & (0.10) & (0.08) & (0.08) & (0.06) & (0.03) & ($<$0.01) & ($<$0.01) \\
		& MM & -3.04 & 1.42 & 3.25 & 4.12 & 4.56 & 4.58 & 4.46 & 4.41 \\
		&& (0.17) & (0.07) & (0.06) & (0.06) & (0.06) & (0.06) & (0.07) & (0.07) \\
		\hline
		\multirow{6}{*}{InvMS}& HEVP & -2.20 & -2.22 & -0.87 & 1.36 & 3.02 & 3.47 & 3.95 & 4.05 \\
		&& (0.03) & (0.06) & (0.11) & (0.06) & (0.03) & (0.03) & (0.02) & (0.02) \\
		& SB & -1.20 & 0.08 & 0.86 & 1.27 & 1.53 & 1.39 & 0.48 & -0.03 \\
		&& (0.01) & (0.03) & (0.03) & (0.03) & (0.02) & (0.01) & ($<$0.01) & ($<$0.01) \\
		& MM & -1.98 & -0.56 & 0.17 & 0.64 & 0.88 & 0.99 & 1.47 & 1.68 \\
		&& (0.03) & (0.05) & (0.05) & (0.05) & (0.06) & (0.13) & (0.27) & (0.29) \\
		\hline
		\multirow{6}{*}{MAX}&HEVP & -0.49 & 2.27 & 3.50 & 4.30 & 4.90 & 5.02 & 5.12 & 5.13 \\
		&& (0.14) & (0.13) & (0.12) & (0.11) & (0.11) & (0.11) & (0.11) & (0.11) \\ 
		& SB & -3.49 & -0.69 & 0.55 & 1.46 & 2.58 & 2.99 & 3.54 & 3.77 \\ 
		&& (0.03) & (0.03) & (0.03) & (0.03) & (0.05) & (0.09) & (0.02) & (0.01) \\
		& MM & -4.99 & -2.42 & -1.05 & 0.17 & 1.21 & 1.53 & 1.46 & 1.41 \\ 
		&& (0.06) & (0.05) & (0.04) & (0.05) & (0.13) & (0.15) & (0.11) & (0.12) \\
		\bottomrule
	\end{tabular}
\end{table}


\end{document}